\documentclass[preprint,nofootinbib,amsmath,amssymb,amsfonts,aps,prd]{revtex4-1}

\usepackage{graphicx}
\usepackage{subfigure}
\usepackage[bookmarks=true, pdfstartview={FitH}, bookmarksopen=true, bookmarksopenlevel=1, linktoc=page]{hyperref}
\usepackage{amsthm}
\usepackage{mathrsfs}

\newcommand{\be}{\begin{equation}}
\newcommand{\ee}{\end{equation}}
\newcommand{\ba}{\begin{aligned}}
\newcommand{\ea}{\end{aligned}}

\begin{document}

\pagestyle{myheadings}

\title{Measure and Probability in Cosmology}
\author{Joshua S. Schiffrin}
\email{schiffrin@uchicago.edu}
\author{Robert M. Wald}
\email{rmwa@uchicago.edu}
\affiliation{Enrico Fermi Institute and Department of Physics \\
  The University of Chicago \\
  5620 S. Ellis Ave., Chicago, IL 60637, U.S.A.}
\date{\today}

\begin{abstract}
General relativity has a Hamiltonian formulation, which formally provides a canonical (Liouville) measure on the space of solutions. In ordinary statistical physics, the Liouville measure is used to compute probabilities of macrostates, and it would seem natural to use the similar measure arising in general relativity to compute probabilities in cosmology, such as the probability that the universe underwent an era of inflation. Indeed, a number of authors have used the restriction of this measure to the space of homogeneous and isotropic universes with scalar field matter (minisuperspace)---namely, the Gibbons-Hawking-Stewart measure---to make arguments about the likelihood of inflation. We argue here that there are at least four major difficulties with using the measure of general relativity to make probability arguments in cosmology: (1) Equilibration does not occur on cosmological length scales. (2) Even in the minisuperspace case, the measure of phase space is infinite and the computation of probabilities depends very strongly on how the infinity is regulated. (3) The inhomogeneous degrees of freedom must be taken into account (we illustrate how) even if one is interested only in universes that are very nearly homogeneous. The measure depends upon how the infinite number of degrees of freedom are truncated, and how one defines ``nearly homogeneous.'' (4) In a universe where the second law of thermodynamics holds, one cannot make use of our knowledge of the present state of the universe to ``retrodict'' the likelihood of past conditions.
\end{abstract}

\maketitle

\tableofcontents

\newpage

\section{Introduction}

Some of the most fundamental issues in cosmology concern the state of
the universe at its earliest moments, for which we have very little
direct observational evidence. In this situation, it is natural to
attempt to make probabilistic arguments to assess the plausibility of
various possible scenarios. For example, one could try to argue that it is
``highly improbable'' that the universe could have been in a
homogeneous, isotropic, and very nearly spatially flat initial state,
so that a mechanism like inflation occurring in the very early
universe is necessary to explain the observed homogeneity, isotropy,
and long age of the present universe. Similarly, one could try to
make arguments
as to whether the occurrence of inflation itself in the early universe
was ``probable.''  Clearly, in order to make any probabilistic
arguments, one needs a measure on the state space of the
universe. General relativity has a Hamiltonian formulation,
so---formally, at least---it gives rise to a canonical measure on
phase space. When restricted to minisuperspace, this canonical 
measure---known as the Gibbons-Hawking-Stewart (GHS) 
measure---has been used to make probabilistic arguments in cosmology, such
as whether inflation was probable \cite{GHS,
HP, GT, CT}. The main purpose of this paper is to provide
critical examination of the use of these types of probability arguments in
cosmology.

As we shall discuss in detail in section \ref{grmeasure},
the phase space of general relativity has many similarities to finite
dimensional phase spaces of classical particle systems, most 
importantly the existence of a
symplectic form which is conserved under time evolution \cite{LW}.
However, there are also significant differences. General relativity is
a field theory, so its phase space is infinite dimensional. In the
finite dimensional case, one can get the conserved canonical
(Liouville) measure by taking the top exterior product
$\wedge^n\omega$ of the conserved symplectic form $\omega$. But in the
infinite dimensional case this does not make sense. Thus, to get a
canonical measure for general relativity, we must truncate the
degrees of freedom by imposing long and
short wavelength cutoffs, so that we are
effectively reduced to finitely many degrees of freedom. A long wavelength
cutoff is naturally imposed by assuming that the cosmological models under
consideration have compact Cauchy surfaces. A short wavelength cutoff 
is effectively imposed by quantum theory and/or the weak coupling of the
degrees of freedom of interest to very high energy modes. With the imposition
of these cutoffs, a canonical measure can be obtained from $\omega$ as in
the finite degrees of freedom case.

However, an even more significant difference between general relativity
and ordinary Hamiltonian systems of particles concerns the
presence of constraints and the nature of these constraints. On account of
the Hamiltonian and momentum constraints, not all points of phase space
are allowed initial data for solutions. Thus, one must restrict consideration
to the submanifold, $\mathcal C$, of points of phase space that satisfy
the constraints. However, when pulled back to $\mathcal C$, the symplectic
form becomes degenerate. The degeneracy directions of the pullback, $\overline{\omega}$,  
of the symplectic form, $\omega$, 
to $\mathcal C$ are precisely the infinitesimal gauge 
transformations, i.e., the changes in initial data corresponding to performing
infinitesimal spacetime diffeomorphisms on the corresponding solutions. One can
remove this degeneracy in a natural manner by passing to the gauge equivalence classes of
initial data, i.e., by considering a new phase space whose points correspond to
physically distinct spacetime solutions. One can then define a canonical measure on (a
truncated version of) this space. However this gives rise to a major, additional
conceptual issue: ``Gauge'' (= spacetime diffeomorphisms) includes ``time evolution,''
so once ``gauge'' has been eliminated from the phase space, there is no longer
any notion of ``time evolution.''
But in the absence of nontrivial time evolution, one cannot make 
any arguments concerning dynamical behavior, such as 
equilibration, the second law of thermodynamics, etc.
Note that this is a classical version of the ``problem of time'' that arises
in canonical quantum gravity.

This issue can be dealt with by choosing a representative of each gauge orbit
rather than passing to the space of gauge orbits.  ``Time evolution'' then gets replaced by 
``change of description'' when a different representative of each gauge orbit is chosen.
The main difficulty with proceeding in this manner 
is that it does not appear possible to choose a unique gauge representative globally in 
phase space in a continuous manner, so this strategy 
cannot be implemented globally. Nevertheless, 
it is plausible that Cauchy surfaces
with constant mean curvature $K=K_*$ are unique in certain classes of ``nearly FLRW'' models
with compact Cauchy surfaces.
For the purposes of this paper, we will assume that the condition $K=K_*$ picks out
a unique Cauchy surface in the spacetimes under consideration here. In that case,
``time evolution'' is replaced by ``change of description'' of initial data when the choice of
representative Cauchy surface is changed from
from $K=K_*$ to $K=K'_*$. Thus, $K$ effectively plays the role of ``time'' \cite{York,MarsdenTipler}, and one
recovers a notion of dynamical evolution, although with a ``time dependent'' Hamiltonian.

The above considerations bring general relativity 
to a framework that is close to the framework of
ordinary statistical physics. In ordinary statistical physics, the canonical measure
has been successfully used to make probability arguments, and one might attempt to
make similar arguments in cosmology. However, there are a number of serious
difficulties in doing so. As we shall discuss in section \ref{probinterp},
one difficulty concerns the effective time dependence
of the Hamiltonian and the insufficient
time for equilibration to have occurred on scales comparable to the Hubble radius,
thereby removing the usual justification for the use of the canonical measure.
Another serious problem concerns the fact that even in minisuperspace models, the
phase space of general relativity is non-compact, and the total measure of phase
space is infinite. Thus, the ``probability'' of any subset of phase space is either $0$
(if its measure is finite), $1$ (if the measure of its complement is finite), or ambiguous
(if both the set and its complement have infinite measure). In section \ref{inflationsection}, we shall
discuss the calculation of the ``probability of inflation'' and---in agreement with a much earlier
discussion by Hawking and Page \cite{HP}---explain why different authors have
obtained widely different results using ostensibly the same 
(conserved!) GHS measure on minisuperspace.
In section \ref{perturbations}, we consider inhomogeneous degrees of freedom. We briefly discuss
some of the additional difficulties that arise when considering the inhomogeneous degrees of
freedom in cosmology, such as placing a ``short wavelength cutoff'' in an expanding universe. Most
of section \ref{perturbations} is devoted to 
deriving the modifications of the GHS
measure that take the effects of inhomogeneous perturbations into
account. 
Finally, in section \ref{retro} we discuss the fact that in a universe with a
thermodynamic arrow of time, one cannot ``retrodict'' 
the past by knowing the present state of the universe and
determining which histories are most ``probable'' in
the canonical measure; the second law of thermodynamics guarantees
that the \emph{actual} history of a system is extremely rare in the
space of all possible histories that are consistent with present
course-grained observations. 

We conclude this introduction by mentioning an issue that we will {\em not} 
address in this paper. The fact that we are conscious observers undoubtedly places a bias on our
observations: We are only able to directly observe portions of spacetime that can admit
the presence of conscious observers. Therefore, in order to obtain the probability that a
region of spacetime will be observed to have a certain property, we should multiply
the a priori probability of the existence of such
a region of spacetime by the probability that conscious observers
exist in this region. Unfortunately, there are few things we know less about in science
than the conditions needed for the existence of conscious observers (or, indeed, even
what ``conscious observers''
are). In many discussions, it is implicitly or explicitly
assumed that various conditions we observe to occur in
our universe are 
necessary for conscious observers to exist. It therefore should not be surprising if the
conclusion then drawn is that the universe should look similar to what we observe
it to be, even if
the a priori probability of what we observe (without taking into account
the bias of our existence) is extremely small. Thus, although such anthropic arguments can
be used to attempt to salvage a theory that predicts a very small a priori probability for what we
observe, we do not feel that any reliable anthropic types
of arguments can be made until we have a much better
understanding of the likelihood that conscious observers could exist in conditions very different from
those of our own universe. 
Thus, we shall not consider anthropic arguments in this paper, nor shall we consider
any issues that involve the likelihood of the existence of observers and/or civilizations
with given characteristics \cite{Susskind,Bousso,Page,Linde}, including the ``measure problem'' as discussed in, e.g., \cite{Vilenkin,Guth}.

\section{The Liouville Measure for General Relativity} \label{grmeasure}

\subsection{The Liouville Measure for Particle Mechanics} \label{particlemechanics}

We begin by briefly reviewing how the Liouville measure on phase space
is constructed in the case of usual Hamiltonian particle mechanics. 
For a Hamiltonian system with $n$ degrees of freedom, the phase space
is $2n$ dimensional and can be labeled by the canonically conjugate coordinates
$(q^i,p_i)$. If the Hamiltonian of the system is
$\mathcal{H}(q^i,p_i)$, then the dynamical trajectories of the system in
phase space are the integral curves of the time
evolution vector field $T$ given by 
\be 
T=
\sum_{i=1}^n\left(\frac{\partial\mathcal{H}}{\partial
    p_i}\frac{\partial}{\partial q^i} -
  \frac{\partial\mathcal{H}}{\partial q^i}\frac{\partial}{\partial
    p_i}\right) \, .  
\ee 
The symplectic form on phase space is given by
\be 
\label{omegaparticlemechanics} 
\omega = \sum_{i=1}^n dp_i\wedge dq^i  \, .  
\ee 
It is obvious from its definition that $d \omega = 0$ and that 
$d {\mathcal H} = - T \cdot \omega$, where the ``$\cdot$'' denotes 
the contraction
of $T$ into the first index of $\omega$. It then follows
immediately that 
\be 
\label{Liouvilleomega} 
\mathcal{L}_T \omega =  T \cdot d \omega + d(T \cdot \omega) 
= -d^2 {\mathcal H} = 0 \, .
\ee 
By taking the wedge product of $\omega$ with itself $n$
times one obtains a volume element on phase space known as the
Liouville measure $\Omega$: 
\be \label{Omegafromomega} 
\Omega \equiv
\frac{(-1)^{n(n-1)/2}}{n!} \wedge^n \omega = d^n p \, d^n q  \, .  
\ee
It follows immediately from \eqref{Liouvilleomega} 
that phase space volumes in the Liouville
measure are preserved under time evolution, 
\be 
\mathcal{L}_T \Omega = 0 \, .
\ee 

One may also derive
the symplectic form (and hence, the Liouville
measure) starting with a Lagrangian formulation of the theory, and we will
use an analog of this procedure to 
get the symplectic form in general
relativity. We consider the space of paths $q(t)$. For a Lagrangian of the form
$L(q^i,\dot q^i)$, its first variation under a variation of path from
$q(t)$ to $q(t) + \delta q(t)$ is given by
\be 
\label{deltaLparticle} 
\delta L = \sum_{i}E_i \delta q^i +
\frac{d}{dt}\left( \sum_{i} p_i \delta q^i \right) \, , 
\ee 
where 
\be E_i
\equiv \frac{\partial L}{\partial q^i} - \frac{d}{dt}\frac{\partial
  L}{\partial \dot q^i}
\ee 
are the Euler-Lagrange equations of motion and 
\be 
p_i \equiv \frac{\partial L}{\partial \dot q^i} \, .  
\ee 
We write
\be 
\theta = \sum_{i} p_i \delta q^i 
\label{theta1}
\ee 
and define $\omega[\delta_1 q, \delta_2 q]$ to be the antisymmetrized
second variation of $\theta$,
\be 
\label{omegawithdeltas} 
\omega[\delta_1 q, \delta_2 q] = \delta_1
\left(\theta[\delta_2 q]\right) - \delta_2 \left(\theta[\delta_1
  q]\right)  \, .
\ee 
Here $\delta_1 q (t)$ and $\delta_2 q (t)$ denote two independent 
variations of the unperturbed path $q(t)$. From \eqref{theta1}, we obtain
\be 
\label{omegawithdeltas2} 
\omega[\delta_1 q, \delta_2 q] = \sum_i (\delta_1 p_i)( \delta_2 q^i)
-(1 \leftrightarrow 2) \, .  
\ee 
Thus, at time $t$, $\omega$ depends
on the path variation only via the variation of $q^i$ and $p_i$ at
time $t$. If we define phase space to be the space of all $(q^i, p_i)$,
then $\omega$ is non-degenerate on this space. The quantity
$\omega[\delta_1 q, \delta_2 q]$ at time $t$ corresponds to the symplectic form 
appearing in \eqref{omegaparticlemechanics} contracted into the tangent
vectors in phase space corresponding to the path variations
$(\delta_1 q, \delta_2 q)$ at time $t$.

\subsection{The Phase Space of General Relativity} \label{grphasespace}

We now consider the phase space of general relativity. There are two
major differences in the construction of phase space and the
symplectic form on phase space as compared with the particle mechanics
case: (i) There are constraints in phase space. (ii) The phase space
of general relativity is infinite dimensional.

We will restrict
consideration to globally hyperbolic spacetimes with compact Cauchy
surfaces, so that integrals over Cauchy surfaces will automatically converge. 
We will explicitly consider the case where the matter content is a
minimally coupled scalar field in a potential---since our primary
application will be to address probability issues in inflationary
cosmology---but the discussion would be qualitatively the same for
other matter fields provided only that the complete theory can be derived
from a diffeomorphism covariant Lagrangian. The dynamical fields
are then the spacetime metric $g_{ab}$ and the scalar field $\varphi$,
and we denote them together as $\chi$:
\be
\chi \equiv \left( g_{ab},\varphi \right) \, .
\ee
The Lagrangian 4-form is (using $8 \pi G = m_{pl}^{-2} \equiv 1$ and
the metric signature $(-,+,+,+)$)
\be \label{L}
\mathscr{L}_{abcd} \left[ \chi \right] = \left( \frac{1}{2}R - \frac{1}{2}\nabla^e \varphi \nabla_e \varphi - V(\varphi) \right) \epsilon_{abcd} \, ,
\ee
where $R$ is the scalar curvature of $g_{ab}$, and $\epsilon_{abcd}$
is the volume form associated with $g_{ab}$. The scalar field's
stress-energy tensor is
\be \label{Tab}
T_{ab} = \nabla_a \varphi \nabla_b \varphi - g_{ab} \left(\frac{1}{2} \nabla^c \varphi \nabla_c \varphi + V(\varphi) \right) \, .
\ee

We follow the prescription of \cite{LW} (see also \cite{IW}), to construct the symplectic form on the phase space of this Lagrangian field theory. When we vary the fields, $\delta\chi = \left( \delta g_{ab},\delta \varphi \right)$, the resulting variation of the Lagrangian is
\be \label{dL}
\delta\mathscr{L}_{abcd} \left[ \chi;\delta\chi \right] = \left( E^{ef}\delta g_{ef} + E_{\varphi}\delta\varphi \right) \epsilon_{abcd} + \left( d\theta \right)_{abcd} \, ,
\ee
where
\be
E^{ab} \equiv -\frac{1}{2}\left(G^{ab} - T^{ab}\right)
\ee
and
\be
E_{\varphi} \equiv - \left( \nabla^a \nabla_a \varphi - \frac{dV}{d\varphi} \right) \, ,
\ee
so that $E^{ab} = 0$ and $E_{\varphi} = 0$ are the field equations for $\chi$. The symplectic potential 3-form appearing in \eqref{dL} is given by
\be \label{theta}
\theta_{bcd} \left[ \chi; \delta\chi \right] = \frac{1}{2} \biggl( \nabla_e \delta g^{ae} - \nabla^{a}\delta g - 2 \left( \nabla^{a}\varphi \right) \delta\varphi \biggr) \epsilon_{abcd} \, ,
\ee
where all indices are raised by $g^{ab}$. In parallel with \eqref{omegawithdeltas}, 
for any two field variations $\delta_1 \chi$ and $\delta_2 \chi$, we define 
\be
\omega_{bcd} \left[ \chi ; \delta_1 \chi, \delta_2 \chi \right] \equiv \delta_1 \left(\theta_{bcd} \left[ \chi;\delta_2 \chi \right]\right)  - \delta_2 \left(\theta_{bcd} \left[ \chi;\delta_1 \chi \right]\right) \,  .
\ee
It follows from taking an antisymmetrized second variation of \eqref{dL}
that if $\delta_1 \chi$ and $\delta_2 \chi$
are solutions of the linearized field equations, then
\be
(d \omega)_{abcd} = 0 \, .
\label{dw}
\ee

Performing the variations on \eqref{theta}, we obtain
\be
\ba
 \omega_{bcd} = \frac{1}{2} \Biggl\{& \frac{1}{2}(\delta_1 g^{ae})(\nabla_e \delta_2 g) - (\delta_1 g_{ef})(\nabla^e \delta_2 g^{af}) + \frac{1}{2}(\delta_1 g_{ef})(\nabla^a \delta_2 g^{ef}) \\ 
&- \frac{1}{2}(\delta_1 g)( \nabla^a \delta_2 g)  + \frac{1}{2}(\delta_1 g)( \nabla_e \delta_2 g^{ae}) + 2(\nabla_e \varphi)(\delta_1 g^{ae})(\delta_2 \varphi) \\
&- (\nabla^a \varphi)(\delta_1 g)(\delta_2 \varphi) - 2 (\nabla^a \delta_1 \varphi)(\delta_2 \varphi) \Biggr\} \epsilon_{abcd} - \left( 1 \leftrightarrow 2 \right) \, .
\ea
\ee
We define $\omega \left[ \chi ; \delta_1 \chi, \delta_2 \chi \right] $ by integrating $\omega_{bcd}$ over a Cauchy surface $\Sigma$,
\be
\label{omdef}
\omega \left[ \chi ; \delta_1 \chi, \delta_2 \chi \right] \equiv \int_{\Sigma}\omega_{bcd}  \, .
\ee
It follows immediately from \eqref{dw} that if $\delta_1 \chi$ and $\delta_2 \chi$
are linearized solutions, then $\omega$ is independent of choice of $\Sigma$.
If we work in coordinates $x^\mu$ where $\Sigma$ is a surface of constant $x^0$, then we obtain,
\be \label{omega}
\ba
\omega = \frac{1}{2} \int_{\Sigma}d^3 x \sqrt{-g} \Biggl\{& \frac{1}{2}(\delta_1 g^{0\mu})(\nabla_{\mu} \delta_2 g) - (\delta_1 g_{\mu \nu})(\nabla^{\mu} \delta_2 g^{0 \nu}) + \frac{1}{2}(\delta_1 g_{\mu \nu})(\nabla^0 \delta_2 g^{\mu \nu}) \\ 
& - \frac{1}{2}(\delta_1 g) (\nabla^0 \delta_2 g)  + \frac{1}{2}(\delta_1 g)(\nabla_{\mu} \delta_2 g^{0 \mu}) + 2(\nabla_\mu \varphi)(\delta_1 g^{0\mu})(\delta_2 \varphi) \\
&- (\nabla^0 \varphi)(\delta_1 g)(\delta_2 \varphi) - 2 (\nabla^0 \delta_1 \varphi)(\delta_2 \varphi) \Biggr\}- \left( 1 \leftrightarrow 2 \right) \,  .
\ea
\ee
This can  be rewritten as
\be \label{omegacanonical}
\omega=\int_{\Sigma} d^3x \left\{\frac{1}{2}(\delta_1 \pi^{ab}) (\delta_2 h_{ab}) + (\delta_1 \pi_\varphi)( \delta_2 \varphi) \right\} - (1\leftrightarrow 2) \, ,
\ee
where $h_{ab}$ is the spatial metric on $\Sigma$ induced by $g_{ab}$, the conjugate momentum $\pi^{ab}$ is given by
\be
\pi^{ab}=\sqrt{h}\left(K^{ab}-Kh^{ab}\right)
\ee 
(where $K^{ab}$ is the extrinsic curvature of $\Sigma$), and $\pi_\varphi$ is given by
\be
\pi_\varphi = \frac{\partial \mathscr{L}}{\partial (\nabla_0 \varphi)} = -\sqrt{-g}\,\nabla^0 \varphi \, .
\ee
When written in the form \eqref{omegacanonical}, it is clear 
that at ``time'' $\Sigma$, $\omega$
depends on the field variations $\delta_1 \chi$ and $\delta_2 \chi$ only through the variations of
$h_{ab}$, $\pi^{ab}$, $\varphi$, and 
$\pi_\varphi$ on $\Sigma$. Thus, if we define phase space, 
$\mathcal P$, to
be the space of all fields 
$(h_{ab}, \pi^{ab}, \varphi, \pi_\varphi)$ on $\Sigma$---where 
$h_{ab}$ is a Riemannian metric, 
$\pi^{ab}$ is a tensor density, $\varphi$ is a scalar field, 
and $\pi_\varphi$ is a 
scalar density---then $\omega$ is non-degenerate on $\mathcal P$
and defines a symplectic form.

Up to this point, our construction of the phase space of 
general relativity has been in direct parallel to
the particle mechanics case, with the only significant 
difference being that $\mathcal P$ is infinite dimensional. 
However, a new issue now arises
from the fact that general relativity has constraints---i.e., not 
all points of phase space are dynamically accessible---and we
are only interested in the points in phase space that satisfy the constraints.
However, the pullback, $\overline\omega$, of $\omega$ to the 
constraint surface, $\mathcal{C}$, is degenerate. 
In fact, the degeneracy directions 
of $\overline\omega$ are exactly the ``pure gauge'' variations: 
$\delta\chi$ is a degeneracy direction of $\overline\omega$ if and 
only if $\delta\chi$ corresponds to a field 
variation arising from an infinitesimal spacetime diffeomorphism, 
i.e., $\delta\chi=\mathcal{L}_\xi \chi$ 
for some vector field $\xi^a$. If we wish to obtain a volume
measure on phase space from $\overline\omega$, it 
is essential that $\overline\omega$ be non-degenerate. We can make 
$\overline\omega$ non-degenerate by 
``factoring out'' the degeneracy directions, i.e., by passing
to the space, $\widetilde{\mathcal P}$,
of orbits of $(h_{ab}, \pi^{ab}, \varphi, \pi_\varphi)$ under 
spacetime diffeomorphisms. In other words,
the  ``physical phase space,'' $\widetilde{\mathcal P}$, of 
general relativity can be taken to be
the space of gauge orbits on $\mathcal{C}$, 
and $\overline\omega$ induces a nondegenerate 
symplectic form on this space. However, since
``time evolution'' is implemented by spacetime diffeomorphisms, 
once we have passed to the space of gauge orbits, there 
is no remaining dynamics.
Each point in the physical phase space represents one spacetime, 
once and for all, and there is no
``motion through phase space'' as there was in the particle mechanics case. 
Thus, a form of Liouville's theorem holds for the physical 
phase space of general relativity (for any choice
of measure), but it is
entirely trivial: Volumes are preserved under time evolution, since there is no such thing as time evolution! Clearly, this form of Liouville's theorem will not be useful for making statistical physics arguments.

However, a nontrivial notion of ``time evolution'' can be reintroduced as follows. Instead of passing to the space of gauge orbits, we can instead try to choose a representative of each gauge orbit. Ideally, one would
wish to find a surface, $\mathcal S$, in $\mathcal C$ with the property 
that each gauge orbit in $\mathcal C$ intersects $\mathcal S$
once and only once. The space $\mathcal S$---with symplectic form given by the
pullback of $\overline \omega$
to $\mathcal S$---could then be used to represent the physical phase space. One could then ask how the 
description of the physical phase space would change if one made a different gauge choice,
i.e., if one used a different surface $\mathcal S'$. This ``change of description'' would correspond 
to ``time evolution'' (i.e., change of the choice of ``representative Cauchy surface'' in spacetime for each
solution) together with additional changes in description
induced by spatial diffeomorphisms. Note that, by the statement below \eqref{omdef}, 
the map taking $({\mathcal S}, \overline \omega)$ to
$({\mathcal S}', \overline \omega)$ that associates points lying on the same gauge orbit preserves the
symplectic structure. Thus, the ``change of description'' map has
the same basic character as ordinary time evolution in Hamiltonian mechanics.

Unfortunately, however, it clearly will not
be possible to globally choose 
a smooth surface $\mathcal S$ in $\mathcal C$ with the property that each gauge
orbit intersects $\mathcal S$ once and only once; to do so would be tantamount to giving a 
prescription to uniquely fix coordinates for arbitrary solutions. 
The best one could realistically hope for is to find a family of surfaces $\{{\mathcal S}_\alpha \}$
that  work in localized regions of $\mathcal C$, i.e., 
such that each gauge orbit intersects at least one
${\mathcal S}_\alpha$ and no gauge orbit intersects any individual ${\mathcal S}_\alpha$ more than
once. For spacetimes with a compact Cauchy surface (as are being considered here), the 
choice of representative time slice given by setting the trace, $K$, of the extrinsic curvature equal to a 
given constant, $K_*$, is an excellent candidate for defining a surface, ${\mathcal S}_\alpha$,
that has the desired property for a wide class of trajectories in $\mathcal C$. Specifically,
if the strong energy conditions holds, Brill and Flaherty \cite{BF} have proven uniqueness of a constant mean
curvature foliation, and existence is known to hold in a wide class of models \cite{Bartnik}, although
counterexamples are also known \cite{Bartnik}; see \cite{Rendall,Rendall2} for further discussion.
However, we cannot directly apply any of these results here because the models we consider 
in this paper with a scalar field in a potential do not satisfy
the strong energy condition\footnote{\footnotesize Indeed, satisfaction of the strong 
energy condition would preclude the possibility of inflation.}. 
Nevertheless, for FLRW models (see the next subsection), existence 
of constant mean curvature slices is obvious---they
are the time slices picked out by the homogeneous and isotropic symmetry---and $K$ ($=3H$, 
where $H$ is the Hubble constant) evolves via
\be \label{FE2}
\dot H = -\frac{1}{2}\dot \varphi^2 +\frac{\kappa}{a^2} \, .
\ee
Thus, $H$ is monotonic if the spatial curvature $\kappa$ is $0$ or $-1$, although it need not be
monotonic for $\kappa = +1$. 
It appears plausible 
that solutions that are sufficiently close to FLRW models with $\kappa = 0, -1$ can be uniquely 
foliated by Cauchy surfaces of constant mean curvature, $K$. 
Our main interest in this paper
is in cosmological models that are close to $\kappa=0$ FLRW models, and we will assume
that in the region of $\mathcal C$ under consideration here,
the equation $K=K_*$ (where $K_*$ is any positive constant) uniquely
selects a Cauchy surface for each solution.
For simplicity, we will also assume\footnote{\footnotesize This assumption is not necessary, since instead of 
fixing spatial
coordinates, we could work with the space of orbits under spatial diffeomorphisms.} that 
a prescription has been given to uniquely fix spatial
coordinates on the surface $K=K_*$. 

The ``time evolution'' discussed above
then corresponds to how the initial data $(h_{ab}, \pi^{ab}, \newline \varphi, \pi_\varphi)$ changes
when we go from the Cauchy surface $K=K_*$ to the Cauchy surface $K={K'_*}$.
Thus, $K$ effectively plays the role of ``time'' (sometimes called ``York time'' \cite{York,MarsdenTipler}).
This effective time evolution is generated by the Hamiltonian
\be
\mathcal{H}=\mathcal{N} C+\mathcal{N}^a C_a \, ,
\ee
where $C$ is the Hamiltonian constraint and $C_a$ is the momentum constraint (see, e.g., \cite{WaldGR}):
\be
C=\frac{1}{2}\sqrt{h} \left[ -^{(3)}R + h^{-1} \pi^{ab}\pi_{ab} - \frac{1}{2}h^{-1}\left(\pi^a_{\phantom{a}a}\right)^2 \right]
\ee
\be
C^a =  -\sqrt{h}\, D_b\left( h^{-1/2}\pi^{ab}\right) \, .
\ee
Here the lapse, $\mathcal N$, and the shift, $\mathcal N^a$ must be chosen so as to
preserve the gauge conditions; specifically, the lapse must be chosen so as to maintain the spatial
constancy of
$K$ and the shift must be chosen to preserve whatever spatial coordinate conditions
have been imposed. Thus, $\mathcal N$ and $\mathcal N^a$ will depend upon $(h_{ab}, \pi^{ab}, \varphi, \pi_\varphi)$,
and their dependence on these variables will, in general, be non-local in space.
Since $\mathcal N$ and $\mathcal N^a$ as well as  
$C$ and $C_a$ depend on $K_*$, the Hamiltonian evolution defined
by $\mathcal H$ is analogous to ordinary time evolution with a 
{\em time dependent} Hamiltonian. Note that $\mathcal H = 0$ 
when evaluated on the constraint surface $\mathcal C$ in $\mathcal{P}$,
but its gradient in directions off of $\mathcal{C}$ is nonzero, so it generates
non-trivial ``time evolution'' in the above sense.

The above considerations provide us with a notion of time evolution in general relativity
on a phase space
with a conserved symplectic form. However, the phase space of general relativity is
infinite dimensional, and it does not make mathematical sense
to attempt take an ``infinite wedge product'' analogous to  \eqref{Omegafromomega}
to define a volume measure on phase space. We will deal with this difficulty by simply
assuming that the effective
number of degrees of freedom is actually finite. Since we are, in any case, considering
only spacetimes with compact Cauchy surfaces, there are no degrees of freedom 
corresponding to ``arbitrarily
long wavelengths,'' so the assumption of effectively having 
only finitely many degrees of freedom 
amounts to imposing a short wavelength cutoff on modes. Such a cutoff 
in the degrees of freedom may be expected to arise at the Planck scale due to a discreteness of space or
other fundamental differences between the true theory of nature and 
classical general relativity. However,
an effective cutoff will occur in any case well before the Planck scale due
to the fact that modes of the metric and 
scalar field that are of sufficiently ``high energy'' will 
not get excited\footnote{\footnotesize In classical theory, all modes share energy equally
in equilibrium (equipartition), leading to an ``ultraviolet catastrophe''
if one has infinitely many degrees of freedom. The effective cutoff occurs
because this is not the case in quantum theory.} and can thereby be ignored.
If we consider only finitely many degrees of freedom and neglect any coupling
to the degrees of freedom that are being ignored, then we may define
a conserved Liouville measure via \eqref{Omegafromomega}, in close analogy
to particle mechanics.

In section \ref{perturbations} we will briefly discuss some of the difficulties arising from
performing such a truncation of the degrees of freedom of general relativity.
We will then analyze the effects of the (large number of)
degrees of freedom that are not being truncated on the measure of 
``nearly FLRW'' universes. However, in the next subsection, we will consider
the measure obtained on phase space by the above construction when one
truncates {\em all} of the degrees of freedom except for the homogeneous 
modes (``minisuperspace'').
This will illustrate our general construction in a simple, concrete case
and will also enable a comparison with previous work \cite{GHS,HP,GT,CT}.

\subsection{The Liouville Measure on Minisuperspace} \label{GHSmeasure}

We now consider Friedmann--Lema\^{\i}tre--Robertson--Walker (FLRW) metrics
\be
ds^2 = -dt^2 + a^2 (t) \gamma_{ij}dx^i dx^j \,  ,
\ee
where $\gamma_{ij}$ is a time independent 
Riemannian metric on a 3-space of constant curvature:
\be \label{gamma}
^{(3)}R_{ijkl} = \kappa \left( \gamma_{ik}\gamma_{jl}-\gamma_{il}\gamma_{jk} \right) \, ,
\ee
with $\kappa=-1,0,+1$ for hyperbolic, flat, and 
spherical 3-geometries, respectively. 
As above (see \eqref{L}), the matter consists of a homogeneous scalar field $\varphi(t)$ 
in the potential $V(\varphi)$. 
Inserting two perturbations towards other FLRW models
\be
\ba
\delta g_{ab}&=2a(t)\delta a(t)\gamma_{ab}
\\ \delta \varphi &= \delta\varphi(t)
\ea
\ee 
into \eqref{omega} yields
\be
\omega = \int d^3x \sqrt{\gamma} \left\{ -6 a (\delta_1 \dot a) (\delta_2 a) +3\dot\varphi a^2 (\delta_1 a)(\delta_2 \varphi)+ a^3 (\delta_1 \dot\varphi) (\delta_2\varphi)\right\} - (1\leftrightarrow 2) \, .
\ee
This can be rewritten as
\be
\omega = \int d^3x \sqrt{\gamma} \left\{ (\delta_1 p_a)(\delta_2 a) + (\delta_1 p_\varphi)(\delta_2 \varphi) \right\} - (1\leftrightarrow 2)
\ee
where we have defined
\be \label{momenta}
\ba
p_a &= -6 a\dot{a} \\
p_{\varphi} &= a^3\dot{\varphi} \, .
\ea
\ee
Thus, the initial data of the previous
subsection---consisting of the tensor fields
$(h_{ab}, \pi^{ab}, \varphi, \pi_\varphi)$ on $\Sigma$---now reduces 
simply to the quantities $(a, p_a, \varphi, p_{\varphi})$, which depend
only on $t$. Thus, when reduced to minisuperspace, the infinite dimensional 
phase
space $\mathcal P$ of the previous subsection becomes $4$-dimensional.
Note that $\omega$ is well defined only when the volume of space, 
$a^3\int d^3x \sqrt{\gamma}$, is finite, i.e., only for the case of 
a compact Cauchy surface, as we have been assuming.
Thus, for $\kappa=0$, we consider only the case where
the universe is a 3-torus; for $\kappa=-1$, 
we consider only cases where the universe is compactified
similarly. For the remainder
of the paper we will assume that the spatial coordinates have been 
scaled so that $\int d^3x \sqrt{\gamma}=1$. This enables us to write
\be
\omega = \left\{ (\delta_1 p_a)(\delta_2 a) + (\delta_1 p_\varphi)(\delta_2 \varphi) \right\} - (1\leftrightarrow 2) \, .
\label{omegamini}
\ee
Note that even in the case $\kappa=0$, $a$ has a well defined physical meaning:
$a^3$ is the total volume of space.
Comparing \eqref{omegamini} to \eqref{omegaparticlemechanics} and
\eqref{omegawithdeltas2}, we see that the symplectic form on minisuperspace is
given by
\be
\omega = dp_a \wedge da + dp_\varphi \wedge d\phi \, ,
\label{omegamini2}
\ee
in agreement\footnote{\footnotesize Gibbons, Hawking, and Stewart derived their expression 
for the symplectic form by starting with a Lagrangian 
for $a$ and $\varphi$ which gave the correct FLRW equations of motion.
(That Lagrangian agrees with what would be obtained by plugging an FLRW 
metric into the Lagrangian \eqref{L}, up to total derivatives 
and a factor of the coordinate volume of space.)} 
with Gibbons, Hawking, and Stewart (GHS) \cite{GHS}.

The next step in deriving the measure on minisuperspace 
is to pull back $\omega$ to the constraint surface to get $\overline\omega$. 
In the homogeneous case, there is only one constraint, the 
Hamiltonian constraint, usually written in the form
\be \label{FE1}
H^2 = \frac{1}{3} \left[ \frac{1}{2} \dot{\varphi}^2 + V(\varphi) \right] -\frac{\kappa}{a^2} \, ,
\ee
where $H \equiv \dot{a}/a$ is the Hubble `constant'. Substituting from
\eqref{momenta}, we obtain
\be \label{Hconstraint}
\frac{p_a^2}{6 a} - \frac{p_{\varphi}^2}{a^3} + 6\kappa a -2 a^3 V(\varphi) = 0 \, .
\ee
Since the constraint surface is $3$-dimensional (and $3$ is an odd number),
the pullback,
$\overline \omega$, of the symplectic form to the constraint surface
is automatically degenerate. However, in accord with the ideas of the previous
subsection, we will obtain a non-degenerate symplectic form if we
pull $\overline \omega$ back to a surface $\mathcal S$ that intersects
each ``gauge orbit'' (i.e., dynamical trajectory) exactly once. 
We choose $\mathcal S$ to be the
surface $H=H_*$; this is guaranteed to work for $\kappa = 0, -1$,
but need not define a unique slice for $\kappa = +1$, as noted above (see \eqref{FE2}). 
Since $\mathcal S$ is $2$-dimensional, the Liouville
measure on $\mathcal S$ is just the pullback of $\omega$ to $\mathcal S$,
\be
\Omega_{\textsc{ghs}} = \omega\bigr|_\mathcal S \, .
\ee 

In the case of a massive scalar field
\be
V(\varphi)=\frac{1}{2}m^2\varphi^2 \, ,
\ee
we can write the constraint in the form
\be \label{quadric}
\dot\varphi^2 + m^2 \varphi^2 - 6\kappa \alpha^2 = 6 H^2 \, ,
\ee
where we have defined 
\be
\alpha \equiv 1/a \, . 
\label{alpha}
\ee
The surface $\mathcal S$ is then the 2-dimensional surface in the space $(m\varphi,\dot\varphi,\sqrt{6}\alpha)$ defined by setting $H=H_*$ in \eqref{quadric}. For $\kappa=-1$ this surface is a hemisphere (since $\alpha>0$) of radius $\sqrt{6}H_*$, for $\kappa=0$ it is a ($\alpha>0$ half) cylinder of radius $\sqrt{6}H_*$ centered on the $\alpha$ axis, and for $\kappa=+1$ it is a ($\alpha>0$ half) one-sheet hyperboloid of minimum radius $\sqrt{6}H_*$ centered on the $\alpha$ axis. 

It is convenient to use the coordinates
$(\alpha,\varphi)$ on $\mathcal S$. These coordinates do not properly cover $\mathcal S$, since,
as can be seen from 
\eqref{quadric}, each allowed value of $(\alpha,\varphi)$ corresponds to two points of 
$\mathcal S$, one for each of the two possible signs of $\dot\varphi$---except for the zero
measure set defined by $\dot\varphi=0$. The coordinates are non-singular on each of
the two non-overlapping
patches with $\dot\varphi>0$ and $\dot\varphi<0$, but become singular where $\dot\varphi=0$.
In either of these nonsingular coordinate patches, the pullback of the symplectic form can be written
\be \label{GHSsymplectic}
\omega\bigr|_{\mathcal S} = \frac{-6 \dot H(\alpha,\varphi,H_*)}{\alpha^4 \dot\varphi(\alpha,\varphi,H_*)} d\varphi \wedge d\alpha \, ,
\ee
where $\dot\varphi(\alpha,\varphi,H_*)$ is determined by \eqref{quadric} (with $H=H_*$) up to a sign depending on which coordinate patch the point under consideration lies in, and $\dot H(\alpha,\varphi,H_*)$ is given by \eqref{FE2}. 

For $\kappa=0,-1$, $\dot{H}$ is always negative, 
and we therefore can write the GHS measure as the positive measure on $\mathcal S$
(switching back from $\alpha$ to $a$)
\be \label{OmegaGHS}
\Omega_{\textsc{ghs}} = \frac{-6 a^2 \dot H(a,\varphi,H_*)}{\left| \dot\varphi(a,\varphi,H_*) \right|} d\varphi da =\left( \frac{3 a^2 \left( 6H_*^2- m^2 \varphi^2+4\kappa a^{-2}\right)}{\sqrt{6H_*^2-m^2 \varphi^2+6\kappa a^{-2}}} \right) d\varphi da   \, ,
\ee
in agreement with expressions\footnote{\footnotesize The divergence at $\dot\varphi\rightarrow 0$ is simply due to the breakdown of
our coordinates there. If we instead had used coordinates $(\alpha,\dot\varphi)$, then we would 
have obtained $\omega\bigr|_{\mathcal S} = \frac{-6 \dot H}{\alpha^4 m^2 \varphi} d\alpha \wedge d\dot\varphi$, 
which is well behaved at $\dot\varphi=0$ (but is singular at $\varphi=0$). 
However, in the case $\kappa=0$ we have $\omega\bigr|_{\mathcal S}=0$ at the points where $\dot\varphi=0$. This \emph{is} a real effect, reflecting the 
fact that the trajectories passing through these points are tangent to $\mathcal S$, as they have $\dot{H}=-\frac{1}{2}\dot\varphi^2=0$, from \eqref{FE2}.} given in \cite{GT,CT}.
Liouville's theorem then says that the
measure of any set of trajectories (universes) is independent of the
choice of $H_*$, as can be verified explicitly by using the
evolution equations.

For $\kappa=+1$, \eqref{OmegaGHS} still holds, but the numerator now can be negative in some regions.
This is a direct manifestation of the fact that the surface $\mathcal S$ given by $H = H_*$ does not
have the property that each trajectory intersects $\mathcal S$ only once. In effect, the GHS measure assigns negative weight to the trajectories
that cross $\mathcal S$ in the ``wrong'' direction $\left(\dot{H} > 0\right)$.
However, we can salvage a version of Liouville's theorem in this case by considering 
only the measure of subsets $U \subset \mathcal S$ that include all trajectory crossings, i.e.,
subsets $U$ with the 
property that if $p \in U$ then every intersection of the trajectory determined by $p$ with $\mathcal S$
also lies in $U$. If we consider only trajectories that satisfy $H \rightarrow  \infty$ in the 
asymptotic past and $H \rightarrow -\infty$ in the asymptotic future \cite{HP,BGKZ}, 
then the total number of 
crossings of $\mathcal S$ will be odd, and the measure of any subset $U$ of the above form
will be positive. This measure will be independent of the choice of $H_*$. 

Returning to the case of general $\kappa$, it is easily seen that if we integrate 
\eqref{OmegaGHS} over the allowed range of 
$\varphi$ and $a$, we obtain
\be
\int_{\mathcal S} \Omega_{\textsc{ghs}} = \infty \, ,
\label{infmeas}
\ee
as has been previously noted by a number of authors \cite{HP, GT, CT}. 
Thus, even for this simplest
minisuperspace case, the total volume of the physical phase space is infinite.
In section \ref{inflationsection}, we will discuss the resulting difficulties that arise if one
attempts to make probability arguments using a measure where the measure of
the total space is infinite.

\section{Difficulties with Equilibration} \label{probinterp}

For ordinary systems in statistical physics---such as a box
of gas---the Liouville measure is used
to determine the probability distribution for observables of interest.
More precisely, for a time independent system with $n$ degrees of freedom,
the Liouville measure $\Omega = \wedge^n \omega$ gives rise to a measure 
$\Omega_M$ (the ``microcanonical ensemble'')
on the constant energy surface $\mathcal{H}=E_0$ by the formula
${\Omega} = \Omega_M \wedge d\mathcal{H}$. This has been
successfully used to compute probabilities when the system has energy $E_0$.
Despite assertions sometimes made to the contrary, the 
justification for using this measure is not based 
upon ``ignorance''---nothing can be derived from ignorance---but rather on
the dynamical evolution properties of the system. For a 
suitably ergodic system\footnote{\footnotesize The precise condition needed for the
validity of Birkhoff's theorem is
metric
indecomposability, i.e.,
the constant energy surface cannot be 
written as the union of two sets of positive measure, each of which is
taken into itself under dynamical evolution.}, Birkhoff's theorem
asserts that for almost all states, the fraction of time spent in 
a region of the constant energy surface is proportional
to the $\Omega_M$-measure of that region. Thus, if we
let $\mathcal O$ be the region of the constant energy surface corresponding to
a particular macroscopic property 
of the system, then if
one examines
the system at a ``random time'' during its dynamical history, the
probability that the system will possess the given macroscopic property 
should be proportional to
the $\Omega_M$-measure of $\mathcal O$. 

Of course, the amount of time
needed by the system to properly explore all portions
of its surface of constant
energy is of order the Poincar\'{e} recurrence time, which is much longer than 
any timescale relevant to human observations. Nevertheless, there are
much shorter ``equilibration timescales'' that should yield sufficient time
for the system to explore enough of its surface of constant
energy to justify the use of the Liouville measure to calculate probabilities.
For example, for a box of gas, there is a ``collision timescale''
governing the amount of time needed for a given particle to significantly
change its momentum and a ``diffusion timescale'' governing the amount of
time needed for a given particle to move across the box. If one waits a
time much longer than either of these timescales---but much shorter
than a Poincar\'{e} recurrence time---then the system should have evolved
sufficiently to ``forget'' any ``special'' initial conditions, and the use of 
the Liouville measure to calculate probabilities should be justified.

From the above discussion, it is evident that use of the Liouville
measure to calculate probabilities will {\em not} be justified in the
following (non-exclusive) circumstances: (i) the system is not ergodic, i.e., it does
not explore all of its surface of constant energy; (ii) one has not
waited a time much greater than the equilibration time after the
system was prepared; (iii) the system has a time dependent Hamiltonian that
is varying on a timescale that is small or comparable to the
equilibration time. 

A situation in which case (i) occurs is a system where there are
additional conserved quantities besides energy. However, 
in many cases, this difficulty can be
dealt with by simply adjoining the 
additional conserved quantities to energy as
additional ``state parameters.'' If the system is ergodic on the
level surfaces of all of the state parameters, then probability
arguments similar to those where the system is ergodic on the constant
energy surfaces can be made. However, a much more problematic 
occurrence of case (i) arises if the
system is such that the surfaces of constant energy are non-compact and have infinite $\Omega_M$-measure,
since then the system will not be able to suitably explore all of its available
phase space. An example of such a system is a spatially unconfined
gas. Another example is a gas of point particles confined by a box but
interacting via Newtonian gravity.  If there is no restriction on how
close the particles can approach each other, the surfaces of constant
energy will have infinite $\Omega_M$-measure, since the potential energy can decrease
without bound, allowing the kinetic energy (and therefore the momenta)
to increase without bound. The system will dynamically evolve
in such a way that groups of particles become more and more tightly bound
and the overall system becomes hotter and hotter, but 
``equilibrium'' will never be achieved and
one will not be
able to use statistical physics arguments to 
predict the details of the evolution.  

In case
(ii), the system will ``remember'' its initial state, so the
probabilities for observation of a macroscopic property corresponding
to region $\mathcal O$ may depend as much or more on the manner of
preparation as on the $\Omega_M$-measure of $\mathcal O$. Case
(iii) is essentially a variant of case (ii), but in case (iii) it will
not help to ``wait longer,'' since even the energy of the system at a
late time $t$ will continue to depend on the initial state and the details of the
history of how the Hamiltonian varied with time.

From the discussion of the previous section, it is clear that general relativity
manifests all three of the above characteristics. Even in the minisuperspace case,
the physical phase space of general relativity is non-compact and its total measure
is infinite, so it is manifestly impossible for any solution to suitably ``explore
its phase space.'' Thus, one is automatically in case (i). Furthermore,  
the timescale for dynamical variation of inhomogeneous degrees of freedom
corresponding to modes of
wavelength comparable to the Hubble radius is of order of the age of the universe,
so, clearly, there is insufficient time for equilibration of such degrees of freedom (case (ii) above). 
Indeed, even for subsystems
that are much smaller than the Hubble radius but still large enough
that self-gravitation is important, the ``equilibration time'' will be much
larger than the age of the universe: As remarked above, for
point particles interacting via Newtonian gravity, the phase space has
infinite volume and no equilibrium is ever achieved. General
relativity effectively provides a cutoff to this behavior via the
formation of black holes, so it may be possible for an isolated, confined
system to achieve equilibrium. Nevertheless, the timescale for matter to
form and/or fall into black holes (and for the black holes to merge)
will be very large, so even if case (i) does not apply, we are in the situation of case (ii).
Finally, as we have seen in subsection \ref{grphasespace}, 
the Hamiltonian of general relativity depends upon the ``time,'' $K_*$,
and thus is effectively time dependent with respect to the degrees of freedom whose dynamics
are sensitive to the Hubble expansion. Thus, the very long wavelength degrees of
freedom are in case (iii).

In summary, for isolated subsystems whose size is much smaller than the
Hubble radius and for which self-gravitation is not important, there
is no reason why ergodic-like behavior should not occur. If such subsystems have
been provided sufficient time to equilibrate,
one should expect the usual 
arguments of statistical physics to apply, and the Liouville measure should 
provide probabilities for observing macrostates. However, to determine a probability 
distribution for the macroscopic state of degrees of freedom whose scale is comparable to
the Hubble radius, one will need to know both the probability distribution for initial conditions and 
the evolutionary history of the universe. Even for subsystems that are much smaller than the Hubble
radius, if self-gravitation is important in the dynamics, the equilibration time is so long that
initial conditions will not be ``erased.''

{\em Thus, we claim that the only way to justify the use of the Liouville
measure in cosmology would be to postulate that the initial conditions
of the universe were chosen at random from a probability distribution
given by the Liouville measure.} In other words, one must postulate
a ``blindfolded creator'' \cite{HW} throwing darts
toward a board of initial conditions for the universe that was 
built using the Liouville measure. This is a very different justification
of the Liouville measure
from that used in ordinary statistical physics, and has the status of
an unsupported hypothesis.

Nevertheless, for the remainder of this paper, 
we will accept the validity of the Liouville measure.
In the next section, we discuss the 
difficulties/ambiguities in obtaining probabilities
from the Liouville measure associated with the fact that, 
even in minisuperspace models,
the total measure of physical phase space is infinite.

\section{Difficulties Due to Infinite Total Measure; The Probability of Inflation} \label{inflationsection}

We have seen in section \ref{grmeasure} that---after truncation to a finite number of
degrees of freedom---a Liouville measure, $\Omega$,
can be defined on the physical
phase space $\mathcal S$. Let $X$ denote a physical property of interest and
let $A_X \subset \mathcal S$ denote the region of physical phase space that
possesses that property. Then, if it were the case that 
$\Omega({\mathcal S}) < \infty$, one could
assign a probability, $P(X)$, to $X$ via
\be
P(X) = \frac{\Omega(A_X)}{\Omega(\mathcal S)} \, .
\label{prob}
\ee
However, in the actual case where $\Omega({\mathcal S}) = \infty$ 
(see \eqref{infmeas}), the situation
divides into three cases: (1) If $\Omega(A_X) < \infty$, then one can 
unambiguously assign $P(X) = 0$. (2) If 
$\Omega ({\mathcal S} \setminus A_X) < \infty$,
then one can unambiguously assign $P(X) = 1$. (3) If both 
$\Omega(A_X) = \infty$ and $\Omega ({\mathcal S} \setminus A_X) = \infty$,
then $P(X)$ is ambiguous and cannot be assigned a value without specifying
additional information, such as a ``regularization procedure.'' Here,
by a ``regularization procedure'' we mean a nested sequence\footnote{\footnotesize Here we assume that $\Omega$ is ``$\sigma$-finite'' so that $\mathcal{S}$ can be written as a countable union of subsets of finite measure.} of subsets
$\{{\mathcal S}_n\}$ with $\cup_n {\mathcal S}_n = {\mathcal S}$ and
$\Omega({\mathcal S}_n) < \infty$ for all $n$. Given such a sequence 
$\{{\mathcal S}_n\}$, we could attempt to define $P(X)$ by 
\be
P(X) = \lim_{n \rightarrow \infty} 
\frac{\Omega(A_X \cap {\mathcal S}_n)}{\Omega({\mathcal S}_n)} \, .
\label{prob2}
\ee
The difficulty is that in case (3) it is easy to see that 
one can get any answer one wishes for $P(X)$
(including no answer at all, i.e., non-existence of the limit) by
a suitable choice of $\{{\mathcal S}_n \}$.

This situation is well illustrated by the following simple example
(also discussed in \cite{Guth}). Let
$\mathbb N$ denote the set of positive integers and let $\Omega$ be the measure
on $\mathbb N$ that assigns to each subset of $\mathbb N$ the number
of elements of that subset. Let $X$ be the property of ``evenness'', so that $A_X$
is the subset of even integers. What is the probability of an integer being
even? If we try to directly apply \eqref{prob}, we obtain 
$P(X) = \infty/\infty$, which is ambiguous. However, 
we can obtain a well defined result by choosing a suitable ``regularization
procedure.''
For example, if we order the integers by size
as $\{1,2,3,\dots\}$ and let ${\mathcal S}_n$
denote the integers appearing in the
first $n$ terms of this sequence, \eqref{prob2} yields $P(X) = 1/2$.
On the other hand, if we ordered the integers as $\{1,3,2,5,7,4,\dots\}$
and let ${\mathcal S}'_n$ denote the first $n$ terms of this sequence,
\eqref{prob2} yields $P(X) = 1/3$. 
This illustrates how $P(X)$ depends on the choice of regularization.
In this example, one could argue that the ordering $\{1,2,3,\dots\}$
is more ``natural'' than the ordering $\{1,3,2,5,7,4,\dots\}$, so that
the ``correct'' value of $P(X)$ is $1/2$. However, the key point here is 
is that the determination of $P(X)$---if, indeed, it can be ``determined''
at all---requires more input than $\Omega$.

The issue of the ``probability of inflation'' for minisuperspace models 
with the GHS measure (see subsection \ref{GHSmeasure} above) provides an excellent
illustration of this type of ambiguity. Using a natural regularization
procedure, Gibbons and Turok \cite{GT} 
obtained an extremely small probability
that the universe would have undergone $N \gg 1$ $e$-foldings of inflation.
However, using a seemingly equally natural regularization procedure
Carroll and Tam \cite{CT} obtained a probability very close to unity that
the universe would have undergone a large number of $e$-foldings of inflation.
A clear explanation of the origin of this type of discrepancy was
given long ago
by Hawking and Page \cite{HP}. 
However, since their explanation does not appear to be widely
known and since 
similar discrepancies have been noted in other, related contexts
\cite{AS}, \cite{CK}, we shall provide a full discussion here.

We consider the minisuperspace models of section \ref{GHSmeasure} with 
$V(\varphi)=\frac{1}{2} m^2 \varphi^2$ and $\kappa=0$. We assume
that $m \ll 1$ (in units where $8 \pi G = \hbar = c = 1$), since,
otherwise (see footnote \ref{smm} below), 
it will not be possible for (many $e$-foldings of) inflation
to occur.
The equations of motion \eqref{FE1} and \eqref{FE2} then reduce to  
\be \label{FEm}
H^2 = \frac{1}{6} \left( \dot\varphi^2 +m^2 \varphi^2 \right) 
\ee
\be \label{FE2flat}
\dot H = -\frac{1}{2} \dot\varphi^2 \, .
\ee
The equation of motion for the scalar field (which can be derived from the
above two equations) takes the form
\be \label{scalareqn}
\ddot\varphi +3H\dot\varphi +m^2\varphi=0 \, .
\ee
The GHS measure \eqref{OmegaGHS} on the surface $\mathcal S$ given
by $H = H_*$ reduces (up to a numerical factor) to
\be \label{OmegaGHSm}
\Omega_{\textsc{ghs}} = \sqrt{6H_*^2-m^2 \varphi^2 \,} \, a^2 da d\varphi \, .
\ee
The allowed range of $\varphi$ is 
\be
|\varphi(H_*)| \leq \sqrt{6}\frac{H_*}{m} \, ,
\label{phibnd}
\ee
but the allowed range of $a(H_*)$ is $(0,\infty)$, so the total measure 
of $\mathcal S$ is infinite, as previously noted. 

In the context of this model, we pose the following question:
{\em What is the probability, 
$P(N)$, in the GHS measure that the universe
underwent at least $N \gg 1$ $e$-foldings of inflation
during times when $H \leq 1$} (in units where $8 \pi G = \hbar = c = 1$)?

\begin{figure}[h]
\centering
\mbox{\subfigure{\includegraphics[width=0.45\textwidth]{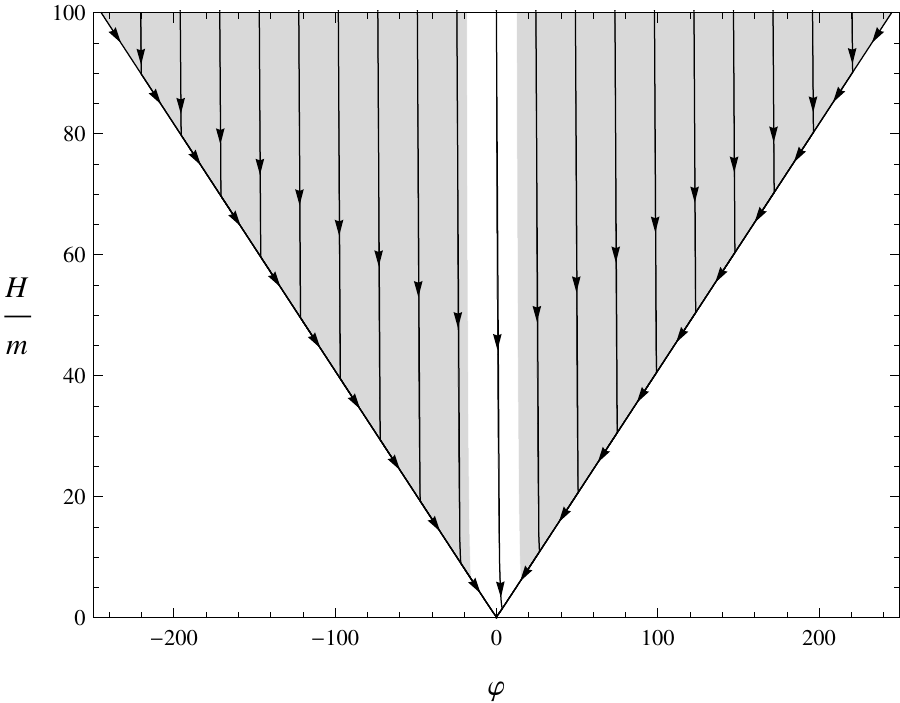}}\qquad\quad 
\subfigure{\includegraphics[width=0.45\textwidth]{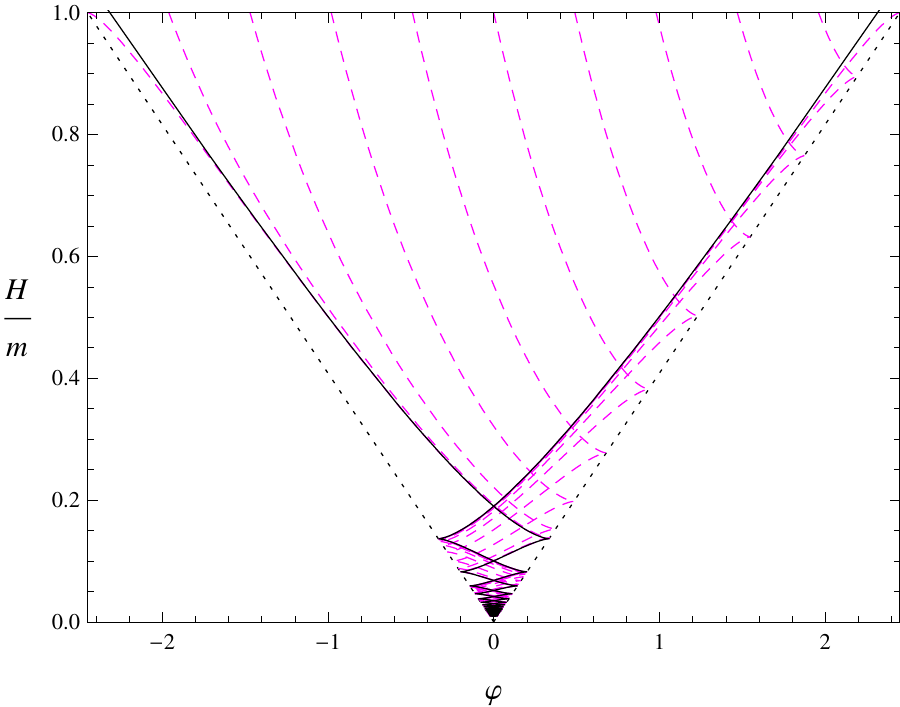} \qquad}}
\caption{\scriptsize The black curves in the left plot are some numerically calculated (flat FLRW-$\varphi$) solutions, chosen so as to be evenly spaced in $\varphi$ at the early time $H=100 m$, and with $\dot\varphi\geq0$ at this early time (the solutions with $\dot \varphi \leq 0$ can be obtained by reflection). Only the behavior of $\varphi$ (horizontal axis) as a function of $H$ (vertical axis) is shown; the behavior of $a$ is not shown, as the $\varphi$ dynamics is independent of the value of $a$. A solution will undergo at least 60 $e$-folds of slow roll inflation if and only if it originates in one of the shaded regions. The right plot is a blow-up of the left plot near the origin. The solid curves are the solutions that have exited slow roll (i.e., they contain all the solutions drawn in the left plot, which are too close together at late times to distinguish on the plot) and the dashed curves are solutions that have never inflated. \\ \hspace*{3mm} For the universes that undergo inflation, we see that $\varphi$ remains practically constant until $\dot\varphi$ reaches to the slow-roll value. Then the solution undergoes slow roll inflation along one of the diagonals (which are close to, but slightly offset from, the turning points $\varphi=\pm\sqrt{6}H/m$) until $H$ becomes small enough ($H_f \sim m$) and inflation ends. If we take the measure surface at an early time (large $H/m$), most of the solutions start out in the shaded region, but if we take the measure surface at a later time, a smaller fraction start out in the shaded region. As we discuss, this does not violate Liouville's theorem, since solutions are ``spreading out'' in the other phase space dimension, $a$.} \label{solutionplots}
\end{figure}

We can estimate the number of $e$-folds of inflation that will occur 
in these models as follows. During slow roll inflation, the potential 
energy dominates the kinetic energy, $\dot\varphi^2  \ll m^2 \varphi^2$, 
and we can approximate \eqref{FEm} and \eqref{scalareqn} by the 
slow roll equations
\be \label{sr1}
H^2 \approx \frac{1}{6} m^2 \varphi^2
\ee
\be 
3H\dot\varphi +m^2\varphi \approx 0 \, ,
\ee
from which it follows that during slow roll we have
\be \label{slowrollvelocity}
\dot\varphi \approx \sqrt{\frac{2}{3}}m \, .
\ee
If inflation occurs, it should end when 
$\dot\varphi^2  \approx m^2 \varphi^2$, so the value of 
$\varphi$ when inflation ends is 
\be
\varphi_f \approx \pm \sqrt{\frac{2}{3}} \, ,
\ee
i.e., inflation ends when $\varphi$ gets too close to the bottom of
the potential\footnote{\footnotesize\label{smm} From \eqref{FEm}, the value of $H$ at the end
  of inflation is $H_f \approx \sqrt{\frac{2}{6}m^2 \varphi_f^2} \sim
  m$, so $m$ must be very small compared to $1$ if
  there is to be any chance of having many $e$-folds of inflation
  after the Planck scale, as claimed above.\label{eoi}}. 
Let $\varphi_i$ denote the value 
of $\varphi$ when $H=1$
\be
\varphi_i \equiv \varphi\bigr|_{H=1}
\ee
At $H=1$, the quantity
$\dot\varphi\left( H=1\right)=\sqrt{6-m^2 \varphi_i^2}$ 
may differ appreciably from the slow roll value
\eqref{slowrollvelocity}. 
Nevertheless, as can be seen from figure
\ref{solutionplots},
if $\varphi_i$ is large enough so that
inflation will eventually occur, then $\dot\varphi$ will approach
the slow roll value before $\varphi$ has a chance to change much,
so the value of $\varphi$ at which
inflation begins is approximately $\varphi_i$. 
Consequently, the number of $e$-folds
of slow roll inflation is approximately\footnote{\footnotesize\label{betterapprox} It is possible to improve this
  approximation. As shown in \cite{GT}, a better approximation to
  the ideal slow roll solution
  is $H=\left(\frac{m\varphi}{\sqrt
      6}\right)\left(1+\frac{1}{3\varphi^2}\right)$.
  Noting that $\dot\varphi=-2
  dH/d\varphi$, we obtain a better approximation for $N$:
\be \nonumber
N \approx \frac{1}{4}\varphi_i^2+\frac{1}{3}\ln\varphi_i \, .
\ee } given by
\be \label{efolds}
\ba
N &\equiv \ln\left(\frac{a_f}{a_i}\right)=\int_{t_i}^{t_f} H dt = \int_{\varphi_i}^{\varphi_f} \frac{H}{\dot\varphi} d\varphi \\
&\approx -\int_{\varphi_i}^{\varphi_f} \frac{\frac{1}{\sqrt{6}}m\varphi}{\sqrt{\frac{2}{3}}m} d\varphi = \frac{1}{4} \left( \varphi_i^2 - \varphi_f^2 \right) \approx \frac{1}{4} \varphi_i^2 \, .
\ea
\ee
Thus, in our approximate treatment of the dynamics,
for $N \gg 1$, the points in phase space for which at least
$N$ $e$-folds of inflation occur at $H \leq 1$ are the ones for
which 
\be \label{condn}
\left|\varphi_i\right| \gtrsim 2\sqrt{N} \, .
\ee 
Whether or not inflation occurs in a given
universe only depends on the initial value of $\varphi$.
In particular, it does not depend on the
initial value of $a$---as one would expect, since the value of $a$ doesn't
affect the dynamics when the spatial curvature vanishes. 

Although the measure of physical phase space is infinite, the range of 
$\varphi$ is bounded by \eqref{phibnd} for any choice of $H_*$, so the infinity
arises entirely from the unbounded range of $a$. Therefore, a 
very simple and natural way of
``regularizing'' the measure is to put in a cutoff, $a_c$, in $a$ and 
then take
the limit as $a_c \rightarrow \infty$.
Since the measure \eqref{OmegaGHSm}
factorizes in $\varphi$ and $a$ 
and the dynamics of $\varphi$ does not depend upon $a$, 
we obtain
\be \label{infratio}
P(N) =\lim_{a_c \rightarrow \infty} \frac{ \, \displaystyle\left(\int_0^{a_c} a^2 da\right) \left( \int_{\varphi_i \ge 2\sqrt{N}} \sqrt{6H_*^2-m^2 \varphi^2 \,}d\varphi \right)\, }{  \, \displaystyle\left(\int_0^{a_c} a^2 da\right) \left( \int_{0}^{\sqrt{6}\frac{H_*}{m}} \sqrt{6H_*^2-m^2 \varphi^2 \,}d\varphi \right)\,} \, ,
\ee
where the $\varphi$-integral in the numerator is taken over the values
of $\varphi$ at $H=H_*$ that come from trajectories with $\varphi \geq
2\sqrt{N}$ at $H=1$. Thus, we can simply cancel the integrals 
over $a$ and---after performing the $\varphi$-integral appearing in the
denominator---we obtain the following well-defined formula for $P(N)$:
\be \label{Pnaive}
P(N) = \frac{2m}{3\pi H_*^2} \int_{\varphi_i \ge 2\sqrt{N}}  \sqrt{6H_*^2-m^2 \varphi^2 \,}d\varphi  \, .
\ee
By Liouville's theorem (see section \ref{grmeasure}), one might expect this expression
to be independent of the choice of $H_*$. 
In fact, however, this expression is not independent
of $H_*$. As we shall now show, if one evaluates $P(N)$ at $H_* \gg m$, 
one obtains a value very close to $1$ (in agreement with Carroll
and Tam \cite{CT}), whereas if one evaluates $P(N)$ at $H_* \sim m$
one obtains a value very close to $0$ 
(in agreement with Gibbons and Turok \cite{GT}).
As we shall then explain, this strong $H_*$-dependence 
of \eqref{Pnaive} arises from the innocuous-looking
regularization we performed in $a$. 

\bigskip

\noindent
{\bf Early-time evaluation ($H_* \gg m$)}: As can be seen from figure
\ref{solutionplots}, $\varphi$ is nearly constant between $H=1$ and $H=H_*$ for trajectories that have not yet started inflation, so \eqref{Pnaive} reduces to
\be \label{earlyP}
\ba
P(N) &\approx  \frac{2m}{3\pi H_*^2} \int_{2\sqrt{N}}^{\sqrt{6}\frac{H_*}{m}}  \sqrt{6H_*^2-m^2 \varphi^2 \,}d\varphi  \\
&= \frac{4}{\pi} \int_{\zeta}^{1} \sqrt{1-x^2}dx \\
&= \frac{2}{\pi}\left[ \cos^{-1}\left(\zeta\right) - \zeta \sqrt{1-\zeta^2} \right] \, ,
\ea
\ee 
where $\zeta \equiv \sqrt{\frac{2 m^2 N}{3 H_*^2}}$. For $\zeta \ll 1$, we can make the approximation
\be
P(N) \approx 1 - \frac{4}{\pi}\zeta \, .
\ee
For the choices $H_* = 1/\sqrt{3}$ and $m = 3 \times 10^{-6}$
considered\footnote{\footnotesize Carroll and Tam quote the value $m=3\times10^{-3}$, but we have confirmed that this was a typo.} by Carroll and Tam  \cite{CT} , this expression yields 
$P(60) \approx 0.99996$, which agrees with their calculation\footnote{\footnotesize Note that for the given values of $H_*$ and $m$, Carroll and Tam calculate numerically that the range of initial values of $\varphi$ which will not give at least 60 $e$-folds of inflation is $-24 \leq \varphi(H_*) \leq 6$ (for the solutions with $\dot\varphi(H_*)>0$), which differs from our estimate \eqref{condn} of $\left| \varphi(H_*) \right| \leq 2\sqrt{60}\approx 15.5$. This discrepancy arises from our approximation that $\varphi$ is constant during the period before inflation; in fact, $|d\varphi/dH|$ is small but not exactly 0, and the solutions will `drift' in $\varphi$ as $H$ changes. However, this drift has negligible effect on the value of $P(N)$, since all not-yet-inflating solutions drift by essentially the same amount (see figure \ref{solutionplots}), and the integrand of \eqref{earlyP} is nearly independent of $\varphi$ near $\varphi=0$.}.
\bigskip

\noindent {\bf Late-time evaluation ($H_* = H_f \sim m$)}: 
This choice of $H_*$ corresponds to the end of inflation for
trajectories that have undergone inflation (see footnote \ref{eoi}).
The slow roll
approximation is no longer valid at this time, so the calculation
can best be done by appealing to Liouville's theorem and evaluating
the phase space volume of interest at $H = 1$. However, in order to apply
Liouville's theorem, we must return to the regulated expression 
\eqref{infratio}, namely
\be
P(N) = \lim_{a_c \rightarrow \infty} \frac{2m}{\pi H_f^2 a_c^3} \int_0^{a_c} a^2 da
\int_{\varphi_i \ge 2\sqrt{N}}  \sqrt{6H_f^2-m^2 \varphi^2 \,}d\varphi  \, .
\ee
The joint integral over $a$ and $\varphi$ is just the GHS measure of
the universes that underwent at least $N$ $e$-foldings of inflation
and have $a \leq a_c$ when $H=H_f$. During the time before a
trajectory starts inflating, the stress-energy tensor of $\varphi$ can
be approximated by that of a perfect fluid with equation of state
$p=\rho$ \cite{HP}. During this period $a \propto t^{1/3}$, so $a
\propto H^{-1/3}$, while $\varphi$ remains practically constant,
$\varphi = \varphi_i$. Once $H$ becomes small enough, slow roll
inflation will begin; by \eqref{sr1}, the value of $H$ when inflation
begins is
\be
H_s \approx \frac{1}{\sqrt{6}} m \varphi_i \, .
\ee 
Using \eqref{efolds}, we then have
\be \label{afracs}
\frac{a\bigr|_{H=H_f}}{a\bigr|_{H=1}} = \frac{a\bigr|_{H=H_f}}{a\bigr|_{H=H_s}} \frac{a\bigr|_{H=H_s}}{a\bigr|_{H=1}} \approx e^{N} H_s^{-1/3} \approx e^{\varphi_i^2/4} \left(\frac{m \varphi_i}{\sqrt{6}}\right)^{-1/3} \, .
\ee 
Thus, the condition $a\bigr|_{H=H_f} \leq a_c$ is approximately equivalent to the condition $a\bigr|_{H=1}<a_c \left(\frac{m \varphi_i}{\sqrt{6}}\right)^{1/3}e^{-\varphi_i^2/4}$. Thus, if we evaluate
the GHS measure of the trajectories that we are considering at $H = 1$,
we obtain (neglecting overall numerical factors of order unity)
\be\ba \label{problate}
P(N) &\propto \lim_{a_c \rightarrow \infty} \frac{1}{m a_c^3} 
\int_{2\sqrt{N}}^{\sqrt{6}/m} d\varphi \sqrt{6-m^2 \varphi^2}
\int_0^{a_c  \left(\frac{m \varphi}{\sqrt{6}}\right)^{1/3}e^{-\varphi^2/4}} a^2 da  \\
&\propto
\int_{2\sqrt{N}}^{\sqrt{6}/m} d\varphi \sqrt{6-m^2 \varphi^2} \, \varphi \, e^{-3\varphi^2/4}
 \\
&\approx 
\int_{2\sqrt{N}}^{\sqrt{6}/m} d\varphi \sqrt{6\,}\,\varphi \, e^{-3\varphi^2/4}  \\
&\propto \left(e^{-3N}-e^{-9m^2/2} \right)  \\
&\approx e^{-3N} \, .
\ea\ee
Note that the factors of $a_c$ canceled in the second line, so the
expression is cutoff independent (and therefore there was no need to
take the limit as $a_c \rightarrow \infty$ after that point).  The
dependence $P(N) \sim \exp(-3N)$ agrees\footnote{\footnotesize Gibbons and Turok 
obtained $P(N)\propto
  e^{-3N}/\sqrt{N}$. Their additional factor of
  $1/\sqrt{N}$ arises from their better estimation of $N(\varphi_i)$
  (see footnote \ref{betterapprox}). 
  If we had used this more accurate formula 
  for $N(\varphi_i)$ in \eqref{afracs}, 
  our calculation would also yield $P(N)\propto
  e^{-3N}/\sqrt{N}$.} with the results of Gibbons and Turok
\cite{GT} and yields $P(N) \sim 10^{-80}$ for $N \sim 60$.

\bigskip

Thus, the situation with regard to obtaining the probability of inflation
in minisuperspace using the GHS measure can now be seen to be closely
analogous to that of
obtaining the probability that an integer is even in the
example given at the beginning of this section. In the integer example, one 
must first ``regularize''
the measure by imposing a ``cutoff'' 
that restricts one to subsets with finitely 
many integers.
Similarly, in the minisuperspace case, one must first 
``regularize'' the measure by putting in a ``cutoff'' in phase space that restricts one to 
subsets of finite volume. In the integer example, a natural cutoff can be
imposed by ordering the integers by size and putting in a cutoff at $n$.
Similarly, in the minisuperspace case, one can order the models by their
``size'', $a$, at ``time'' $H_*$ and put in a cutoff at $a_c$. 
The difficulty is that the subsets of phase space with 
$a \leq a_c$ at ``time'' $H = H_*$ are very different in ``shape'' from the 
subsets of phase space with $a \leq a'_c$ at 
``time'' $H = H'_*$. In particular, the universes with 
$a \leq a_c$ at an early time $H_* \gg m$ that will undergo inflation
will be systematically ``pushed out'' to much larger values of $a$
at a late time $H_* \sim m$ than universes that do not inflate. Therefore
if one regularizes by imposing a cutoff in $a$ at a late time, one will find
that a much smaller percentage of universes will have inflated, as our
calculations above explicitly show.

Should one impose a cutoff in $a$ at, say, the Planck time (see \cite{CK})
and conclude
that inflation is highly probable? Or, should one impose a cutoff in
$a$ at a late time \cite{Turoktalk} and conclude that inflation is highly
improbable? Or, should one impose an entirely 
different regularization scheme and
perhaps draw an entirely different conclusion? Our purpose here is not
to answer these questions but to emphasize that, even in this simple
minisuperspace model, one needs more information
than the GHS measure to obtain the probability of inflation.

\section{Contribution of Inhomogeneous Degrees of Freedom} \label{perturbations}

\begin{figure}[h]
\centering
\mbox{\subfigure{\includegraphics[width=0.46\textwidth]{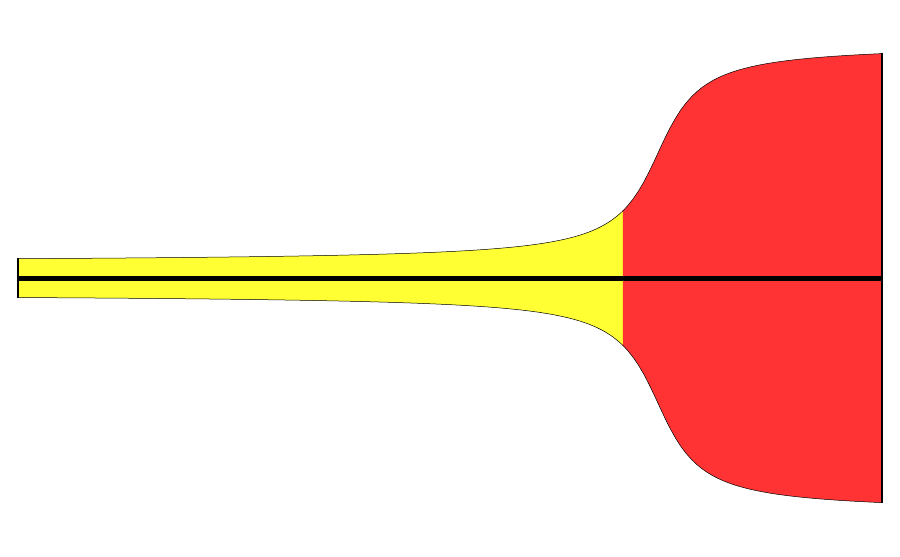}}}
\caption{\scriptsize A region of the plane ${\bf R}^2$. 
The black line $y=0$ lies
  more in the yellow region than in the red region, but the red region
  is larger. In the ``real'' measure, $\Omega=dx\, dy$, a randomly
  chosen point is more likely to lie in the red region. But in the
  restricted measure, $\Omega\bigr|_{y=0}=dx$, a
  randomly chosen point is more likely to lie in the yellow
  region.} \label{shape}
\end{figure}

In the previous section, we restricted our considerations to
minisuperspace, i.e., the two dimensional submanifold of homogeneous,
isotropic (FLRW) solutions. As discussed in section \ref{grmeasure}, the pullback of
the symplectic form to minisuperspace yields the GHS measure, which we
used for the calculations of the previous section. However,
minisuperspace is a set of measure zero in the full phase space. Even
if we are only interested in ``nearly FLRW'' solutions, it is far from
clear that the GHS measure will give a valid estimate of the phase space
measure of the spacetimes that are ``close'' to a given FLRW
solution. This difficulty is well illustrated by the following simple
example (see figure \ref{shape}): Consider the plane ${\bf R}^2$, with
measure $\Omega = dx dy$, consider the line $y=0$ with measure
$\Omega\bigr|_{y=0} = dx$, and consider the (union of the) yellow and red
shaded regions. If we choose a point in the shaded regions of the
plane ``at random'' with respect to the measure $\Omega$, we will find
that it is more likely that the point lies in the red region than the
yellow. But if we try to estimate this probability by choosing a point
on the line ``at random'' with respect to the measure $\Omega\bigr|_{y=0} =
dx$, we will get the wrong answer, i.e., we will find that it is more
likely that it lies in the yellow region. It is quite possible that the
situation for the phase space of general relativity could be quite
similar: Specifically, the plane with measure $\Omega = dx dy$ could be
analogous to the full phase space of general relativity 
with the measure of subsection \ref{grphasespace}; the line $y=0$
with measure $\Omega\bigr|_{y=0} = dx$ could be analogous to minisuperspace
with the GHS measure of subsection \ref{GHSmeasure}; the (union of the) yellow and red shaded regions
could be analogous to the spacetimes that, physically, are ``close'' to
FLRW models; and the yellow and red regions could correspond to different
properties of interest, e.g., the yellow region could correspond to universes
that do not inflate, and the red region could correspond to universes that
do. The minisuperspace calculation would then erroneously 
predict a low probability of inflation,
since the full phase space calculation would give a high probability of inflation.

In this section, we will calculate the effects of including
inhomogeneities on the measure of nearly FLRW models, thereby
obtaining the modifications to the GHS measure produced by the type of
effect illustrated in figure \ref{shape}. Before presenting these
calculations, however, we briefly discuss three significant
difficulties of principle---for which we have no satisfactory
resolutions---regarding the inclusion of inhomogeneous degrees of
freedom, even when we restrict consideration to nearly FLRW models.

\subsection{Difficulties Arising in the Treatment of Inhomogeneous Degrees of Freedom} \label{infdof}

The most serious difficulty arising in the treatment of inhomogeneous
degrees of freedom is that there are infinitely many of them.
As discussed in section \ref{grmeasure}, in order to effectively reduce the
system to a finite dimensional phase space, it is necessary to impose
a short wavelength cutoff. There are obvious difficulties in imposing
such a cutoff, since there is arbitrariness in the choice of cutoff, and,
whatever choice is made, there
undoubtedly will remain some coupling 
between the long wavelength degrees of
freedom that are being kept, and the short wavelength degrees of freedom
that are being discarded. However, in cosmology, there is an additional
major issue of principle that arises from the fact that the physical
wavelength of a mode is time dependent. In particular, if inflation
occurs, a degree of freedom corresponding to a linearized perturbation of wavelength of order of the Planck scale at a time prior to inflation could
correspond to a wavelength of order of the Hubble radius or larger in
the present universe. If one imposes a short wavelength cutoff at a
fixed physical scale such as the Planck scale, then the dimension of
the phase space will vary with time, making it impossible to apply
Liouville's theorem or any usual statistical physics arguments. On
the other hand, if one puts in a cutoff at a fixed co-moving
wavelength so that the number of degrees of freedom does not change
with time, then, in order to consider
physically relevant degrees of freedom in the present universe,
it may be necessary to consider degrees of freedom with wavelength smaller than the Planck length in the very early universe.

Another difficulty is that the notion of whether an 
inhomogeneous cosmology is ``nearly
FLRW'' is a highly time dependent notion. Our own universe is believed
to have been extremely homogeneous and isotropic on all scales during
a much earlier phase, but is highly inhomogeneous on small
scales now. On the other hand, most universes that are nearly
homogeneous and isotropic at a Hubble constant corresponding 
to the present value, $H_0$, in our universe were extremely inhomogeneous
at a much earlier phase \cite{CT}. Thus, the universes that are
``nearly FLRW'' at $H=H_0$ comprise a very different class of universes
than those that are ``nearly FLRW'' at $H \gg H_0$.

Finally, even if we select a particular value, $H=H_*$, to impose a
short wavelength cutoff and impose the requirement that inhomogeneities
be ``small,'' it is not obvious how to define ``smallness.'' We will make 
such a choice in our calculations below, but one could argue
that other choices are equally ``natural,'' and it is clear that
the results will depend in a
significant way upon the definition of ``smallness.''

\subsection{The Symplectic Form for Perturbations of FLRW} \label{scalarperts}

In section \ref{GHSmeasure} we evaluated the symplectic form of
general relativity, $\omega[\chi; \delta_1 \chi, \delta_2 \chi]$, when
$\delta_1\chi$ and $\delta_2\chi$ are \emph{homogeneous} perturbations
off an FLRW background $\chi$; the result of this calculation is given
by \eqref{omegamini2}. In this subsection, we will evaluate
$\omega[\chi; \delta_1 \chi, \delta_2 \chi]$ when
$\delta_1 \chi$ and $\delta_2 \chi$ are arbitrary (inhomogeneous) 
perturbations off of
an FLRW background.  This will enable us to obtain the Liouville
measure on the space of ``nearly-FLRW'' universes, which we shall do
in the next subsection.

We will mostly follow the notation of Mukhanov et al. \cite{Mukhanov} 
(although we use the opposite metric signature). In conformal time,
\be \label{eta}
\eta = \int \frac{dt}{a} \, ,
\ee
the metric of an FLRW universe takes the form
\be \label{rw}
ds^2 = a^2 (\eta) \left(- d\eta^2 + \gamma_{ij}dx^i dx^j \right) \, ,
\ee
and the Friedmann equations take the form
\be \label{background}
\widetilde{H}^2 + \kappa = \frac{1}{3} \left( \frac{1}{2} (\varphi_0 ')^2 + a^2 V(\varphi_0) \right)
\ee
\be
\frac{a''}{a} + \kappa = -\frac{1}{3} \left( \frac{1}{2} (\varphi_0 ')^2 - 2 a^2 V(\varphi_0) \right) \, ,
\ee
where $'$ denotes differentiation with respect to $\eta$, and 
\be \label{conformalH}
\widetilde{H} \equiv \frac{a'}{a} = \dot{a} = a H \, .
\ee
We write, 
\be
\xi_{a} \equiv \nabla_{a} \eta \, ,
\ee
and we define $D_a$ to be the derivative operator associated with
$\gamma_{ab}$.

As is well known, there is a natural way to
decompose any perturbation of an FLRW solution
into its scalar, vector, and
tensor parts \cite{Bardeen, Mukhanov},
\be
\delta g_{ab} = \left(\delta g_{ab}\right)_{\text{scalar}} + \left(\delta g_{ab}\right)_{\text{vector}} + \left(\delta g_{ab}\right)_{\text{tensor}} \, ,
\ee
where
\begin{eqnarray} \label{deltag}
\left(\delta g_{ab}\right)_{\text{scalar}} &=& 2 \, a^2 \left( -\phi \xi_a \xi_b + \xi_{(a}D_{b)}B - \psi \gamma_{ab} + D_a D_b E \right)\\  \label{deltagv}
\left(\delta g_{ab}\right)_{\text{vector}}&=&a^2\left(\xi_{(a} U_{b)}+D_{(a} V_{b)}\right) \\ \label{deltagt}
\left(\delta g_{ab}\right)_{\text{tensor}}&=&a^2 h_{ab} \, .
\end{eqnarray}
Here $U^a$ and $V^a$ are tangent to the spatial slices and are divergence
free ($D_a U^a = D_a V^a = 0$), whereas $h_{ab}$ is also
tangent to the spatial slices, traceless
($h^a_{\phantom{a}a}=0$), and divergence free ($D^a h_{ab}=0$). In
addition to the metric, we must also perturb the scalar field
$\varphi$,
\be \label{deltas}
\varphi(\eta,x^i) = \varphi_0 (\eta)+\delta\varphi(\eta,x^i) \, ,
\ee
and $\delta \varphi$
is classified as belonging to the scalar part of the perturbation. 

It is well known that the 
different perturbation types are decoupled in the linearized
limit, i.e. the linearized Einstein equations break up into
independent scalar, vector, and tensor pieces. The
symplectic form, given by \eqref{omega}, is quadratic in the
perturbations, but it is not difficult to see that the ``cross terms''
between the different perturbation types integrate to zero, so we have
\be 
\omega = \omega_{\text{scalar}}+\omega_{\text{vector}}+\omega_{\text{tensor}} \, ,
\ee
where $\omega_{\text{scalar}}$ only contains terms quadratic in scalar
perturbations, $\omega_{\text{vector}}$ only contains terms quadratic
in vector perturbations, and $\omega_{\text{tensor}}$ only contains
terms quadratic in tensor perturbations. 

We will now calculate the scalar, vector, and tensor perturbation
contributions to the symplectic form, 
$\omega$, and its pullback, $\overline \omega$, to the
constraint submanifold.

\subsubsection{Scalar Perturbations}

Scalar perturbations naturally decompose into a ``homogeneous part''
(corresponding to perturbations towards other FLRW models) and an
``inhomogeneous part'' (for which the spatial integrals of $\phi$, 
$\psi$, and $\delta \varphi$ vanish). The computation of 
$\omega[\chi; \delta_1 \chi, \delta_2 \chi]$, when
$\delta_1\chi$ and $\delta_2\chi$ are homogeneous perturbations
was already done in section \ref{GHSmeasure}. There are no cross terms in $\omega$
between homogeneous and inhomogeneous perturbations, so it remains only
to calculate $\omega[\chi; \delta_1 \chi, \delta_2 \chi]$, when
$\delta_1\chi$ and $\delta_2\chi$ are inhomogeneous perturbations.

From \eqref{deltag} and \eqref{deltas}, we see that there are 5 scalar
functions, $(\phi,\psi,B,E,\delta\varphi)$, defining an inhomogeneous scalar
perturbation. However, there is gauge freedom to eliminate two of
these functions via $\delta g_{ab} \rightarrow \delta g_{ab} +
\nabla_{(a}v_{b)}$ with $v_{a} = \alpha \xi_a + D_a \beta$. Thus,
there are three independent gauge-invariant variables
\cite{Bardeen}, and we will use the choice of variables 
given in \cite{Mukhanov}:
\be \label{GI}
\ba
\Phi &= \phi + \frac{1}{a}\left[a\left(B-E'\right)\right]'   \\
\Psi &= \psi - \frac{a'}{a}(B-E')       \\
\sigma &= \delta\varphi + \varphi_0 ' (B-E') \, .
\ea
\ee
In terms of
these variables, the scalar linearized Einstein's equations are \cite{Mukhanov}
\be \label{hc}
D^2 \Psi - 3 \widetilde{H}\left(\widetilde{H}\Phi + \Psi'\right) +3\kappa\Psi = \frac{1}{2}\left[\varphi_0 '\left(\sigma'-\varphi_0 '\Phi\right) +a^2 \left. \frac{dV}{d\varphi}\right|_{\varphi_0} \sigma\right]
\ee
\be \label {mc}
D_i \bigl( \widetilde{H} \Phi + \Psi' -  \frac{1}{2} \varphi_0 ' \sigma \bigr) = 0
\ee
\be \label{ev}
\ba
\biggl[ \Psi'' + \widetilde{H} \left( \Phi'+2 \Psi' \right) + \left( 2 \widetilde{H}' + \widetilde{H}^2 \right) \Phi  - \kappa \Psi &+ \frac{1}{2} D^2 (\Phi-\Psi) \biggr] \delta^i_{\phantom{i}j}  - \frac{1}{2}D^i D_j (\Phi - \Psi) 
\\ \qquad &=  \frac{1}{2} \left[ \left( \sigma'-\varphi_0 ' \Phi \right) \varphi_0 ' -a^2  \left. \frac{dV}{d\varphi}\right|_{\varphi_0} \sigma \right] \delta^i_{\phantom{i}j} 
\ea
\ee
where spatial indices are raised by $\gamma^{ij}$, and $D^2 \equiv D^i D_i$. Equation \eqref{ev} implies that
\be \label{phipsi}
\Phi = \Psi \,  .
\ee

Plugging an FLRW solution and two inhomogeneous perturbations of the form
(\ref{deltag}, \ref{deltas}) into the formula, \eqref{omega}, for
$\omega$, and using \eqref{GI} and \eqref{phipsi}, we obtain
\be \label{omegaperts}
\ba
\omega_{\substack{\text{scalar}\\ \text{inhom}}}
= -& 2 a^2 \int_{\Sigma} d^3 x \sqrt{\gamma} \left\{ 3 \Phi_1 ' \Phi_2 +  \frac{1}{2} \left[ 4 \varphi_0 ' \Phi_1 \sigma_2 - \sigma_1 ' \sigma_2 \right]  \right\}- (1 \leftrightarrow 2)
\\  &+ \left( \text{\small terms that vanish when the linearized Einstein
equations are satisfied} \right) \, .
\ea
\ee
Here we have integrated by parts and dropped surface terms, since we
are only considering universes with compact spatial slices. 
We have not explicitly written out the terms that vanish when the linearized
Einstein equations
are satisfied, since for the calculation of the Liouville measure,
we will pull back the symplectic form to the constraint submanifold,
and these terms will vanish. Note that, as we
discussed in section \ref{grmeasure}, the symplectic form should have ``pure
gauge'' variations as its degeneracies when pulled back to the
constraint surface; this can be seen explicitly in \ref{omegaperts}
since, when the linearized Einstein equations are
satisfied, the non-vanishing terms have been 
written exclusively in terms of gauge invariant variables.

Using \eqref{hc}, \eqref{mc}, and \eqref{phipsi} to eliminate $\sigma$ and $\sigma'$ in terms of 
$\Phi$ and $\Phi'$, we obtain
\be \label{obar}
\ba
\omega_{\substack{\text{scalar}\\ \text{inhom}}} =  -&\frac{4 a^2}{(\varphi_0 ')^2} \int_{\Sigma} d^3 x \sqrt{\gamma} \Bigl\{  \Phi_1' \left(D^2+3 \kappa \right) \Phi_2  \Bigr\} -(1\leftrightarrow 2) 
\\  &+ \left( \text{\small terms that vanish when the linearized Einstein
equations are satisfied} \right) \, .
\ea
\ee
We can further rewrite this expression by eliminating $\Phi'$ in terms of the gauge
invariant density contrast \cite{CT}
\be
\delta \equiv \frac{\delta\rho + \rho_0 ' (B-E')}{\rho_0}
\ee
where
\be
\rho \equiv  -\frac{1}{2}\nabla^a \varphi \nabla_a \varphi +V(\varphi) \, .
\ee
Using \eqref{hc}, we find that when the linearized Einstein equations hold,
we have
\be \label{deltaeqn}
\delta = \frac{2}{a^2 \rho_0}\left[ D^2 \Phi - 3 \widetilde{H}\left(\widetilde{H}\Phi + \Phi'\right) +3\kappa\Phi \right] \, .
\ee
We thereby obtain
\be \label{omegaperts2new}
\ba
\omega_{\substack{\text{scalar}\\ \text{inhom}}} &= \frac{2 a \rho_0}{3 H \dot\varphi_0 ^2} \int_{\Sigma} d^3 x \sqrt{\gamma} \Bigl\{  \delta_1 \left(D^2+3 \kappa \right) \Phi_2  \Bigr\} -(1\leftrightarrow 2)
\\  &\qquad+ \left( \text{\small terms that vanish when the linearized Einstein
equations are satisfied} \right) \, ,
\ea
\ee
where we have used \eqref{eta} and \eqref{conformalH} to switch from conformal time to proper time.

We can expand a general inhomogeneous perturbation in eigenfunctions
of the Laplacian $D^2$ (which has a discrete spectrum since our
spatial slices are compact). Let $f_{(n)}(x^i)$ for $n=0,1,2,...$ satisfy
\be \label{ef1}
D^2 f_{(n)} = -k_n^2 f_{(n)}
\ee
with $k_0 = 0$ and $k_n$ nondecreasing in $n$, and
\be \label{ef2}
\int_{\Sigma} d^3 x \sqrt{\gamma} f_{(m)} f_{(n)} = \delta_{n,m} \, ,
\ee
where we remind the reader that we have normalized
$\gamma_{ij}$ so that $\int_{\Sigma} d^3 x\sqrt{\gamma}\equiv 1$. 
We expand
\be \label{expanded}
\ba
\Phi &= \sum_{n=1}^\infty \Phi^{(n)}f_{(n)}
\\ \delta &= \sum_{n=1}^\infty \delta^{(n)}f_{(n)} \, .
\ea
\ee
Substituting these expansions into \eqref{omegaperts2new}, pulling the
expression back to the constraint submanifold, and adding in
the homogeneous contribution $\omega_{\textsc{ghs}}$, we obtain
\be \label{omegabar}
\overline{\omega}_{\text{scalar}} =  \overline\omega_{\textsc{ghs}} + \sum_{n=1}^\infty \overline\omega_{\text{scalar}}^{(n)} \, ,
\ee
where
\be \label{omegaperts2}
\overline\omega_{\text{scalar}}^{(n)} =\frac{ 2 a \rho_0 \left( k_n^2 - 3\kappa \right) }{3 H \dot\varphi_0^2 } d\Phi^{(n)} \wedge d\delta^{(n)} = \frac{ a \left( H^2 + \kappa a^{-2} \right) \left( k_n^2 - 3\kappa \right) }{ H \bigl( 3H^2- V(\varphi_0) +3\kappa a^{-2} \bigr) } d\Phi^{(n)} \wedge d\delta^{(n)} \, .
\ee

Finally, we comment that an irrotational fluid with equation of state
$P = w \rho$ can be treated within our formalism by setting 
$V=0$ and
replacing
$X \equiv (-\frac{1}{2}\nabla^a \varphi \nabla_a \varphi)$ with 
$X^\ell$ in the Lagrangian \eqref{L}, with $\ell =
  \frac{1+w}{2w}$. Carrying out the analogous calculations
in the case $\kappa = 0$, we obtain
\be\label{A1}
\overline\omega_{\text{fluid}}^{(n)} = \frac{2 k_n^2 a \rho_0}{3 H \left( \rho_0 + p_0 \right)}  
d\Phi^{(n)} \wedge d\delta^{(n)} \,  .
\ee
This agrees with the expression obtained by
Carroll and Tam \cite{CT} for the case $p=\rho/3$.

\subsubsection{Vector Perturbations} \label{vectorsection}

From \eqref{deltagv}, we see that there are two divergence-free spatial vectors, $U_i$ and $V_i$, defining a vector perturbation. However, there is gauge freedom to eliminate one of
these vector fields via $\delta g_{ab} \rightarrow \delta g_{ab} +
\nabla_{(a}v_{b)}$ with $v_{a}$ divergence free and tangent to the spatial slices. The combination 
\be
\overline{V}_i \equiv U_i - V_i '
\ee
is gauge invariant \cite{MBook}. One can calculate that the only term appearing in the integrand 
of $\omega_{\text{vector}}$ is proportional to 
$ \left[(V_1)^i D^2 (\overline{V}_2)_i - (1\leftrightarrow 2)\right]$.
However, the constraint equation is \cite{MBook} 
\be
D^2 \overline{V}_i = 0 \, ,
\ee
where the right hand side is zero because the stress energy tensor has no vector part for scalar field matter. Thus, we have
\be
\overline\omega_{\text{vector}}=0 \, .
\ee

\subsubsection{Tensor Perturbations} \label{tensorsection}

The tensor perturbation \eqref{deltagt} involves a symmetric,
traceless, and divergence-free spatial tensor field, $h_{ij}$. There
are no constraints on $h_{ij}$ or its time derivative arising from
Einstein's equation, and $h_{ij}$ is gauge invariant.  Plugging two
tensor perturbations \eqref{deltagt} into the formula \eqref{omega}
for the symplectic form, we obtain
\be
\omega_{\text{tensor}}  = -\frac{1}{4}a^2 \int_\Sigma d^3 x \sqrt{\gamma} \left( (h_1)_{ij} \frac{d(h_2)^{ij}}{d\eta}  - (1\leftrightarrow 2) \right) \, .
\ee

Let $\left(f_{(n)}\right)_{ij}$ for $n=1,2,...$
denote an orthonormal basis of 
divergence free and traceless
eigentensors of $D^2$, i.e., 
\be
D^2 \left(f_{(n)}\right)_{ij} = -\widetilde{k}_n^2 \left(f_{(n)}\right)_{ij} \, ,
\ee
with $\widetilde{k}_n$ nondecreasing in $n$, and
\be
\int_{\Sigma}d^3 x \sqrt{\gamma} \left(f_{(n)}\right)_{ij} \left(f_{(m)}\right)^{ij} = \delta_{n,m} \,
\ee
(we remind the reader that spatial indices are raised by $\gamma^{ij}$). We expand
\be
h_{ij} = \sum_{n=1}^{\infty} h^{(n)}(\eta) \left(f_{(n)}\right)_{ij} \, .
\ee
(Note that we are including both polarizations in the sum.)
We obtain
\be \label{tensoromega}
\omega_{\text{tensor}} = -\frac{1}{4}  \, a^2 \sum_{n=1}^{\infty} dh^{(n)} \wedge d\left(h^{(n)}\right)' = \frac{1}{4}  \, a^3  \sum_{n=1}^{\infty} d\dot{h}^{(n)} \wedge dh^{(n)}\, .
\ee
The formula for $\overline\omega_{\text{tensor}}$ is identical, as there are no tensor constraints.

\subsection{Modifying the GHS Measure by Including Inhomogeneities, and the Effect on the Probability of Inflation} \label{mmm}

As discussed is section \ref{grmeasure}, in order to obtain a finite dimensional phase
space, we impose a short wavelength cutoff. 
We do so by discarding all scalar
modes that have $n>\mathcal{N}_1$ and all tensor modes that have
$n>\mathcal{N}_2$. (Note that if we put the cutoffs at approximately
the same values of wavelength, i.e., if $k^2_{\mathcal{N}_1} \approx
\widetilde{k}^2_{\mathcal{N}_2}$, then we will have $\mathcal{N}_2  \sim
2\mathcal{N}_1$, since there are two tensor polarizations.)
Imposing these cutoffs and combining
\eqref{omegabar} with \eqref{tensoromega}, we obtain 
the following expression for the symplectic form
for perturbations of FLRW, pulled back to the constraint surface:
\be \label{omegabar2}
\overline{\omega} = 
\overline\omega_{\textsc{ghs}} +  \frac{ 2 a \rho_0  }{3 H \dot\varphi_0^2 } \sum_{n=1}^{\mathcal N_1} \left( k_n^2 - 3\kappa \right) d\Phi^{(n)} \wedge d\delta^{(n)} +  \frac{1}{4}  \, a^3 \sum_{n=1}^{\mathcal N_2}   d\dot{h}^{(n)} \wedge dh^{(n)} \, .
\ee
As discussed in section \ref{grmeasure}, to obtain a nondegenerate 
symplectic form, we further restrict
to the surface $\mathcal S$
defined by $K=3H_*$, i.e., we impose $H=H_*$ on the background
FLRW spacetime and $\delta K = 0$ on the perturbations.
In addition, as mentioned in \ref{grphasespace}, we
must fix the spatial coordinates as well. However, 
our variables describing inhomogeneous perturbations
have been written in a gauge invariant form, so there is no change
in their functional form when we restrict to $\delta K = 0$ and
fix spatial coordinates. Furthermore, homogeneous spatial
diffeomorphisms act trivially on the homogeneous perturbations.
Thus, we obtain
\be
\ba
\omega\bigr|_{\mathcal S} &= \left( \frac{3 a^2 \left( 6H_*^2- m^2 \varphi^2+4\kappa a^{-2}\right)}{\sqrt{6H_*^2-m^2 \varphi^2+6\kappa a^{-2}}} \right) d\varphi \wedge da \\ 
&\quad+  \frac{ a \left( H_*^2 + \kappa a^{-2} \right)  }{ H_* \bigl( 3H_*^2- V(\varphi_0) +3\kappa a^{-2} \bigr) } \sum_{n=1}^{\mathcal N_1}\left( k_n^2 - 3\kappa \right)d\Phi^{(n)} \wedge d\delta^{(n)} \\
&\quad+ \frac{1}{4}  \, a^3 \sum_{n=1}^{\mathcal N_2} d\dot{h}^{(n)} \wedge dh^{(n)} \, ,
\ea
\ee
where we have used the explicit form of the GHS measure, \eqref{OmegaGHS}, derived in
section \ref{GHSmeasure}. 

To obtain the Liouville measure on our truncated, $2(1+\mathcal
N_1+\mathcal N_2)$-dimensional phase space, we take the
top exterior product of $\omega\bigr|_{\mathcal S}$, thereby obtaining
\be\label{Omegatilde}
\ba
\Omega \equiv& \frac{1}{(1+\mathcal N_1+\mathcal N_2)!}\wedge^{(1+\mathcal N_1+\mathcal N_2)}\omega\bigr|_{\mathcal S} \\
&=\left( \frac{3 a^2 \left( 6H_*^2- m^2 \varphi^2+4\kappa a^{-2}\right)}{\sqrt{6H_*^2-m^2 \varphi^2+6\kappa a^{-2}}} \right) d\varphi \wedge da \\ 
&\quad \wedge  \left( \frac{ a \left( H_*^2 + \kappa a^{-2} \right)  }{ H_* \bigl( 3H_*^2- V(\varphi_0) +3\kappa a^{-2} \bigr) } \right)^{\mathcal N_1} \prod_{n=1}^{\mathcal N_1}\left( k_n^2 - 3\kappa \right)d\Phi^{(n)} \wedge d\delta^{(n)} \\
&\quad \wedge \left( \frac{1}{4}  \, a^3 \right)^{\mathcal N_2} \prod_{n=0}^{\mathcal N_2}  d\dot{h}^{(n)} \wedge dh^{(n)} \, .
\ea
\ee
This expression for the $2(1+\mathcal N_1+\mathcal N_2)$-form 
$\Omega$ is, of course, valid only when evaluated at exact FLRW models. However,
if we restrict consideration to solutions that are sufficiently close to FLRW models,
we should be able to neglect the dependence of $\Omega$ on  
$(\Phi^{(1)}, \delta^{(1)}, \dots, \Phi^{(\mathcal N_1)}, \delta^{(\mathcal N_1)},
\dot{h}^{(1)} h^{(1)}, \dots, \dot{h}^{(\mathcal N_2)} h^{(\mathcal N_2)})$ and
continue to use \eqref{Omegatilde} to calculate phase space volumes of regions
sufficiently near FLRW models.

We now wish to integrate $\Omega$ over a range of
perturbation variables corresponding to ``small
inhomogeneities,'' so as to obtain the ``true'' measure of nearly FLRW spacetimes (see
figure \ref{shape} and the associated discussion at the beginning of this section).
To do so, we need to decide how to define ``small
inhomogeneities'', i.e., how to choose which universes are ``nearly
FLRW''. Following Carroll and Tam \cite{CT}, we
will choose the variables $\Phi$ and
$\delta$ to indicate the ``size'' of scalar perturbations, 
since $|\Phi|\ll1$ means that the metric perturbation is
small compared to the background metric, and $|\delta|\ll 1$ means
that the density perturbation is small compared to the background
density. For the tensor
perturbations, we will choose $h$ and
$\dot{h}/H$ as indicators of ``size,'' since
$|h|\ll1$ means the metric perturbation is small, and
$|\dot h|\ll H$ means that the perturbation of the extrinsic curvature
is small compared to the background.
Thus, we provisionally define 
``nearly FLRW'' universes at ``time'' $H_*$ to be those that have $|\Phi(H=H_*)|\leq
\epsilon$, $|\delta(H=H_*)|\leq \epsilon$, $|h(H=H_*)|\leq
\epsilon$, and $|\dot h(H=H_*)|\leq H_*\epsilon$ for some
$\epsilon >0$. (Here, for simplicity,
we have taken $\epsilon$ to be independent of $n$ and the same for scalar and tensor
modes.)
Then, the modified GHS measure on minisuperspace, which takes the
``thickening'' effects of inhomogeneities into
account, is given by
\be  \label{modifiedOmega}
\ba
\widetilde{\Omega}_{\textsc{ghs}}  &\equiv \int\limits_{\substack{|\Phi^{(n)}| \leq \epsilon \\ |\delta^{(n)}| \leq \epsilon \\ |h^{(n)}|\leq\epsilon \\ |{\dot h}^{(n)}|\leq  \epsilon H_*\\}}{\Omega} \quad \\
&= \,  \left( \frac{ 4 \epsilon^2 a \left( H_*^2 + \kappa a^{-2} \right) }{ H_* \bigl( 3H_*^2- V(\varphi_0) +3\kappa a^{-2} \bigr) } \right)^{\mathcal N_1} \cdot \left(\prod_{n=1}^{\mathcal N_1} \left( k_n^2 - 3\kappa \right)\right) \cdot \left( \epsilon^2 a^3 H_* \right)^{{\mathcal N_2}} \cdot \Omega_{\textsc{ghs}}\, ,
\ea
\ee
where $\Omega_{\textsc{ghs}}$ is given\footnote{\footnotesize Note that 
$\Omega_{\textsc{ghs}}$ is conserved under dynamical
evolution in minisuperspace, but $\widetilde{\Omega}_{\textsc{ghs}} $ 
is not because the ``boxes'' of size
$\epsilon$ in our perturbation variables do not maintain their 
size under evolution.} by equation \eqref{OmegaGHS}.
For the case of $\kappa=0$ and
$V(\varphi)=\frac{1}{2}m^2\varphi^2$, and choosing $\mathcal N_2=2\mathcal N_1 \equiv 2 \mathcal N$, we obtain
\be \label{modOm}
\widetilde{\Omega}_{\textsc{ghs}}  \propto  \left( \frac{ a^7 H_*^3 }{  6H_*^2- m^2\varphi_0^2  } \right)^{\mathcal N}  \Omega_{\textsc{ghs}}  = \left( \frac{ a^7 H_*^3 }{  6H_*^2- m^2\varphi_0^2  } \right)^{\mathcal N}  \sqrt{6H_*^2-m^2 \varphi_0^2 \,} \, a^2 da d\varphi_0 \, .
\ee
Note that $\widetilde{\Omega}_{\textsc{ghs}} $ varies with $a$ as
$a^{2+7\mathcal N}$, so including more perturbation modes makes
the large-$a$ divergence more severe. In particular, including the effect of
perturbations on the measure does not alleviate the difficulties
discussed in section \ref{inflationsection} associated with the infinite total measure
of minisuperspace.

It is easily seen that $\widetilde{\Omega}_{\textsc{ghs}}$ has a 
non-integrable divergence (for $\mathcal N \geq 2$)
as $\varphi_0 \rightarrow \pm\sqrt{6}H_*/m $, which lies at the boundary of the range of $\varphi_0$ allowed
by the Hamiltonian constraint \eqref{FEm}, and corresponds to $\dot\varphi_0 \rightarrow 0$. The origin of this
divergence can be traced to the fact that $\Phi^{(n)}$ and $\delta^{(n)}$ 
become linearly dependent at $\varphi_0 = \pm\sqrt{6}H_*/m $. Thus, as 
$\varphi_0 \rightarrow \pm\sqrt{6}H_*/m $, a box of ``size'' $\epsilon$
in $\Phi^{(n)}$ and $\delta^{(n)}$ allows an unboundedly large range of other variables such as
 \be \label{sdotblowup}
\dot{\sigma}^{(n)} \approx \frac{m^2 \varphi_0}{H_* \dot\varphi_0^2}\left(\frac{m^2 \varphi_0^2}{6  }\delta^{(n)} + \frac{2 k_n^2 }{3 a^2 }\Phi^{(n)} \right) \, ,
\ee
where $\sigma$ was defined by \eqref{GI}. We will therefore supplement our definition of ``nearly FLRW''
universes by imposing the additional requirement that $|\dot{\sigma}^{(n)}| \ll m |\varphi_0|$
when $\dot{\varphi}_0$ is close to zero. This
eliminates the divergence at $\varphi_0 \rightarrow \pm \sqrt{6}H_*/m$ 
that occurs in \eqref{modOm} without affecting our formula for the measure when
$\varphi_0$ is not close to  $\pm \sqrt{6}H_*/m$.

Let us now compute the probability of inflation in parallel with the analysis of section \ref{inflationsection}, but
using the measure $\widetilde{\Omega}_{\textsc{ghs}}$ rather than $\Omega_{\textsc{ghs}}$.
First, we perform an ``early-time evaluation'' ($H_* \gg m$) of the probability of inflation. 
For fixed $a$, the GHS measure is peaked at $\varphi_0=0$ and decreases to zero as $\varphi_0$ approaches the boundary of its allowed range, so the GHS measure is peaked on universes that have the 
least amount of inflation.
By contrast,  $\widetilde{\Omega}_{\textsc{ghs}}$ 
is more peaked near the boundary of the allowed range of $\varphi_0$, where
the most inflation occurs. Thus, the probability of inflation computed using $\widetilde{\Omega}_{\textsc{ghs}}$ 
at early times will be even higher than the (extremely high) value obtained in section \ref{inflationsection} using $\Omega_{\textsc{ghs}}$.

On the other hand, suppose we perform a 
late-time evaluation ($H_* \sim m$) of the probability of inflation.
In this case, the main effect of including the perturbation modes comes from the change in the $a$ 
dependence of the measure from $\Omega_{\textsc{ghs}} \propto a^2$ to 
$\widetilde{\Omega}_{\textsc{ghs}} \propto a^{2+7\mathcal N}$.
Carrying out the analysis in parallel with section \ref{inflationsection}, we obtain
\be
P(N) \sim e^{-(3+7\mathcal N)N} \, ,
\ee
which is {\em much} smaller\footnote{\footnotesize Indeed, if the ``box size,'' $\epsilon$,
is not extremely small, one may argue that the probability of inflation
should be even further reduced because some of the universes that are deemed
to have undergone inflation in this calculation will not have done so
because the spatial inhomogeneities were too large.} 
than the (extremely small) probability of
inflation calculated in section \ref{inflationsection} using $\Omega_{\textsc{ghs}}$. 

As in section \ref{inflationsection}, our purpose here is not to advocate for performing
the calculations at early times or at late times, nor to advocate for
our particular choice of definition of ``nearly FLRW''
spacetimes. Rather, our purpose is merely to illustrate in a concrete
manner the additional issues that arise---and the additional choices
that need to be made---when one takes the inhomogeneous degrees of
freedom into account.

\section{Impossibility of Retrodiction} \label{retro}

In the previous sections, we have highlighted difficulties that arise
when one attempts to use the natural Liouville measure of general
relativity to calculate probabilities in cosmology, such as the
probability of inflation. Some of the difficulties are a direct
consequence of the fact that the measure of phase space is
infinite. A possible approach to dealing with this problem is to use
our knowledge of (or beliefs about) the present universe to reduce the
relevant region of phase space to one that has finite volume. In
particular, we know/believe that the present universe is nearly an
FLRW model. We do not know the present value of the scalar factor,
$a$, of this background FLRW model, but we can simply assume that it
takes some particular value $a_0$. Suppose we restrict consideration to the
region, $A$, of phase space corresponding to universes that 
differ by a suitably small (but finite)
amount from this background FLRW model at the
time when the trace of the extrinsic curvature is equal to the present value,
$H_0$, of the Hubble constant. If we let $B$ denote the 
subset of universes that undergo suitably many $e$-foldings of inflation,
then one might argue that---taking account of our knowledge of
the present state of the universe---the probability 
that our universe underwent an era of inflation should be
given by
\be 
P = \frac{\Omega(B \cap A)}{\Omega(A)} \, , 
\ee 
where
$\Omega$ is the canonical measure of general relativity, and the right side
is now well defined, since $A$ and $B \cap A$ have finite measure.
Indeed, this formula essentially corresponds to the calculation of 
the previous subsection taking $H_* = H_0$; a calculation of this
nature was also carried out in \cite{CT}. We refer to this type of calculation 
as a ``retrodiction'' because it uses present conditions 
as an input to determine the
likelihood of past conditions. The purpose of this section is to argue
that, in a universe in which the second law of thermodynamics is valid, 
this type of calculation is essentially 
{\it guaranteed} to give the {\it wrong} answer,
i.e., it will assign an extremely low probability to what actually 
did happen in the past. This ``futility of retrodiction'' has been
discussed in depth by Eckhardt \cite{Eckhardt}, and we refer the reader to that reference for additional discussion.

In ordinary statistical mechanics, we can use knowledge about the
present macrostate of a system to successfully
predict the likely future evolution
of the system. This is done by considering all of
the possible microstates that are consistent with the present
macrostate and evolving them forward in time using the (microscopic)
dynamical equations. If it is found that an overwhelming majority (in
the Liouville measure) of these microstates will have some given
property at some specified time in the future, then we can be
confident in predicting that the physical system will have that
property at that specified time. We refer to this type of argument
as a ``trajectory counting argument''. The key point is that although one
can use a trajectory counting argument to predict the future, one
cannot use a trajectory counting argument to ``retrodict'' the
past. The time reversal invariance of the microscopic dynamics assures
us that
most trajectories through phase space
that pass through a present non-equilibrium macrostate have entropy
that is increasing both into the future and into the past. However, 
it appears that we live in a
universe in which the second law of thermodynamics holds, where
entropy increases only in one time direction (``the future'').
In practice, we find that trajectory counting arguments do give correct
results when used to make predictions; i.e., systems do not 
appear to be evolving
towards some ``special final state.''
However, because entropy in our universe decreases 
towards the past,
trajectory counting arguments
will always give incorrect answers when used to make 
retrodictions.

A simple example will illustrate this point. Suppose that at 12:00
there is a cup of water with a temperature slightly above freezing and
with a small ice cube floating in it. It is known that the cup was
isolated since at least 11:59, and it will remain isolated until at
least 12:01. Was there a larger ice cube at 11:59? A trajectory
counting argument for this case would consider all possible histories
that have a small ice cube in the cup at 12:00 and calculate the
fraction (in the Liouville measure) of those histories for which there
is a larger ice cube in the cup at 11:59. Very few trajectories have
this property; the overwhelming majority 
have a smaller ice cube (or only liquid water) at
11:59. Indeed, since this system is time reversal invariant, the
trajectory counting argument will give the same answer whether we go
forward or backward in time. The trajectory counting argument will
correctly predict that there will be less ice by 12:01. It will
incorrectly retrodict that there was less ice at 11:59.

The same behavior applies in cosmology. If, as above, we let $A$ be the set of
universes that can be described as ``nearly-FLRW today,''
then most universes in $A$ were extremely inhomogeneous at early
times \cite{CT}. Indeed, if we consider almost any universe that looks like ours now
and reverse-evolve it into the past, we would find that
inhomogeneities would generically grow (into the past)---so much that
they could no longer be described by the linearized theory. Inflation
(or, ``deflation,'' since we are evolving backwards) would
surely not occur at early times. In other words, in the set of all possible
histories that are consistent with our current conditions, the
most numerous type of history (in the Liouville measure) is one in
which the universe was highly inhomogeneous in the past---with many 
regions of ``delayed big bangs,'' i.e., ``white holes''---and just
managed to smooth itself out enough by our present time so that it
could be approximately described today as a linear perturbation of
FLRW. 

As we have just argued, in a universe in which the second law of
thermodynamics holds, we cannot test a hypothesis about the past by
retrodiction, i.e., we cannot test it by determining whether the
hypothesized occurrence is generic in the space off all possible
histories consistent with present observation. Rather, to ascertain
what happened in the past, we make hypotheses about the past based on
considerations such as simplicity (\`a la ``Occam's Razor'') and
mathematical elegance. Once we have made such a hypothesis, we can
test it using {\em prediction} rather than retrodiction. If a simple
and/or elegant hypothesis about the past makes predictions---via
trajectory counting arguments---that are in agreement with
observation, this is taken as evidence in favor of the hypothesis. The
simpler and more elegant the hypothesis and the more it accounts for,
the stronger we consider the evidence in its favor. In a universe in
which the second law of thermodynamics is obeyed, this is essentially
the {\em only} way to get information about the past state. For
example, we deduce that light element nucleosynthesis must have
occurred in the early universe not by taking the nuclear abundances of
the present universe and retrodicting past de-synthesis, but by making
very simple hypotheses about the state of the early universe and
predicting the abundances we see today. The evidence for light element
nucleosynthesis is strong because much is explained from a simple
hypothesis.

In a similar manner, we assert that
{\em if one wishes to make an argument in favor 
of inflation having
occurred in the early universe, this argument
must be based upon its being a simple
and/or elegant hypothesis that accounts for observed phenomena.} 
Any argument about the ``likelihood'' of inflation based upon
the Liouville (or other) measure on phase space will require 
a justification for the use of this measure (see section \ref{probinterp}) as well
as a justification for the regularizations used to obtain probabilities
from this measure (see sections \ref{inflationsection} and \ref{perturbations}). Furthermore, as we have 
argued in this
section, we cannot obtain any useful information about whether inflation 
occurred in the early universe by determining whether it is generic
in the space of all universes that look like ours today.

\begin{acknowledgments}
This research was supported in part by
NSF Grant No. PHY08-54807 to the University of Chicago.
\end{acknowledgments}

\bibliography{mybib}

\end{document}